\documentclass[trackchanges, ]{aastex701}

\usepackage{amsmath}
\usepackage{multirow}

\submitjournal{ApJ}

\shorttitle{Non-spherical Potential Field Model}
\shortauthors{Wu et al.}

\begin{document}

\title{A Non-Spherical Model for the Solar Coronal Magnetic Field}

\author[orcid=0000-0002-1349-8720,gname=Ziqi,sname=Wu]{Ziqi Wu}
\affiliation{School of Earth and Space Sciences, Peking University, Beijing, China}
\affiliation{Center for mathematical Plasma Astrophysics, Department of Mathematics, KU Leuven, Leuven, Belgium}
\email{wuziqi@pku.edu.cn}  

\author[orcid=0000-0001-8179-417X,gname=Jiansen, sname=He]{Jiansen He} 
\affiliation{School of Earth and Space Sciences, Peking University, Beijing, China}
\affiliation{State Key Laboratory of Solar Activity and Space Weather, National Space Science Center, Chinese Academy of Sciences, Beijing, China}
\email[show]{jshept@pku.edu.cn}
\correspondingauthor{Jiansen He}

\author[orcid=0000-0001-7205-2449, gname=Chuanpeng,sname=Hou]{Chuanpeng Hou}
\affiliation{Institut für Physik und Astronomie, Universität Potsdam, Potsdam, German}
\email{}

\author[orcid=0000-0001-9628-4113,gname=Tom,sname=Van Doorsselaere]{Tom Van Doorsselaere}
\affiliation{Center for mathematical Plasma Astrophysics, Department of Mathematics, KU Leuven, Leuven, Belgium}
\email{}

\author[orcid=0000-0003-4726-9755, gname=Rui, sname=Zhuo]{Rui Zhuo}
\affiliation{School of Earth and Space Sciences, Peking University, Beijing, China}
\affiliation{Center for mathematical Plasma Astrophysics, Department of Mathematics, KU Leuven, Leuven, Belgium}
\email{}

\author[orcid=0000-0001-7170-0408, gname=Tianhang, sname=Chen]{Tianhang Chen}
\affiliation{School of Earth and Space Sciences, Peking University, Beijing, China}
\email{}

\author[orcid=0000-0003-4716-2958, gname=Liping, sname=Yang]{Liping Yang}
\affiliation{State Key Laboratory of Solar Activity and Space Weather, National Space Science Center, Chinese Academy of Sciences, Beijing, China}
\email{}

\author[orcid=0000-0002-1089-9270, gname=David, sname=Pontin]{David Pontin}
\affiliation{University of Newcastle Australia, Callaghan, New South Wales, Australia}
\email{}

\author[orcid=0000-0002-0497-1096, gname=Daniel, sname=Verscharen]{Daniel Verscharen}
\affiliation{Mullard Space Science Laboratory, University College London, London, UK}
\email{}

\author[orcid=0000-0002-4935-6679, gname=Fang, sname=Shen]{Fang Shen}
\affiliation{State Key Laboratory of Solar Activity and Space Weather, National Space Science Center, Chinese Academy of Sciences, Beijing, China}
\email{}

\begin{abstract}
The coronal magnetic field plays a fundamental role in governing coronal activities, driving space-weather events, and shaping the heliosphere. Due to a lack of direct observations, extrapolation models such as the Potential Field Source Surface (PFSS) model become the primary method to obtain the three-dimensional magnetic field distribution in the corona. However, the PFSS model cannot solve the long-standing open-flux problem, in which the extrapolated open magnetic flux is significantly lower than that inferred from in-situ measurements. To address this issue, we develop a Non-Spherical Potential Field (NSPF) model. The model introduces a Non-Spherical Source Surface (NSSS) defined as an isosurface of the total magnetic field. The NSSS naturally forms concave structures beneath external current sheets, enabling the model to generate substantially more open magnetic flux while yielding a physically plausible distribution of open field regions. As a result, the NSPF model successfully reproduces complex coronal magnetic topologies, interplanetary magnetic field properties, and solar wind source mappings. Our refined coronal magnetic model provides a useful framework for future research on solar and heliospheric magnetic coupling.
\end{abstract}


\section{Introduction}
The coronal magnetic field governs the release of the solar wind and drives eruptive events that propagate into the heliosphere. It is therefore a key factor for understanding coronal activity, identifying solar wind magnetic connectivity, and monitoring solar-planetary interactions. However, due to the weak magnetic field strength and high temperature in the corona, it is extremely difficult to measure the magnetic field directly through the Zeeman effect \citep{Babcock1967, Crutcher2019, Schad2024}, which is widely used to obtain the photospheric magnetograms. Despite efforts with other observational techniques, such as coronal seismology \citep{Yang2024}, so far, obtaining coronal magnetic field data with sufficient spatial and temporal resolution remains a major challenge.

As a result, extrapolation models are widely used to infer the coronal magnetic field distribution \citep{Regnier2013, Wiegelmann2017}. The existing models include the magnetohydrodynamic (MHD) models \citep{VDH2014, Mikic2018, Liu2026}, magnetohydrostatic (MHS) models \citep{Zhu2022}, magneto-frictional models \citep{Guo2016A, Guo2016B}, force-free models \citep{Wiegelmann2006, Wiegelmann&Sakurai2021} , outflow equilibrium model \citep{Rice2021}, and potential field models \citep{Altschuler&Newkirk1969, Schatten1969}. 

The Potential Field Source Surface (PFSS) model is the most widely used method to extrapolate coronal magnetic fields \citep{Altschuler&Newkirk1969, Aschwanden2005, Stansby2020}. It relies on two key assumptions: (i) the potential field hypothesis, which assumes that there are no currents in the model, and (ii) the source surface hypothesis, which assumes that there is a spherical surface at a certain height where the magnetic field is radial. The lower boundary (photosphere)takes synoptic magnetograms as input, and the upper boundary (source surface) has zero magnetic potential. Between the photosphere and the source surface, the magnetic field is obtained by solving the Laplace equation $\nabla^2 u=0$ for the magnetic scalar potential $u$. The PFSS model is primarily designed to capture the large-scale, quasi-steady structures of the coronal magnetic field. It is also widely applied in the two-step ballistic backmapping techniques \citep{Gieseler2023, Badman2020}, and commonly utilized as the initial magnetic field distribution in global MHD simulations \citep{Arge2003, Keppens2023}.

To compensate for the limitation introduced by the potential field hypothesis, many extensions of the PFSS model have been developed to include current sheets. \citet{Schatten1971} developed the Schatten Current Sheet (SCS) model—also referred to as PFSS + potential field current sheet (PFCS), where an external shell is added to the PFSS domain. In this shell, the radial field at the source surface is first converted to a single polarity by reversing the sign of the weaker polarity regions, and the Laplace equation is then solved again. This procedure produces a global, thin heliospheric current sheet (HCS) consistent with interplanetary observations \citep{Knizhnik2024, Shi2024}. \citet{ZhaoHoeksema1994, ZhaoHoeksema1995} further extended this idea with the Current Sheet-Source Surface (CSSS) model, which includes horizontal currents in two layers of the SCS configuration \citep{Koskela2019}. These models provide a more realistic description of the global coronal and heliospheric magnetic field \citep{Ma2025}.

Other intrinsic limitations of the PFSS model stem from the source surface hypothesis. All field lines reaching the source surface are assumed to be open and extend radially into interplanetary space. A spherical source surface causes unrealistic field line geometries because it forces all field lines to become radial beyond the source surface. Moreover, the source surface height determines the amount of open magnetic flux and thus influences the inferred source regions of the solar wind, particularly for the slow solar wind that is most likely to originate from open-closed field boundaries \citep{Abbo2016, Wilkins2025}.
A canonical value of 2.5 R$_\odot$ is commonly adopted based on coronal loop observations \citep{Schatten1969}, which is also compatible with the open solar flux calculated directly from the photospheric magnetic field by a vector sum method \citep{Tahtinen2024, Tahtinen2026}. In many cases, however, the source surface height is treated as a free parameter that is chosen such that the model results match in-situ measurements. \citep{Panasenco2020}.
Numerous studies have sought to constrain the optimal source surface height using observations such as eclipse white-light images \citep{Benavitz2024}, coronal hole observations and loop statistics \citep{Heinemann2026}, in-situ interplanetary magnetic field (IMF) measurements \citep{Arden2014, Lee2010, Shoda2025}, combined data sources \citep{Badman2022}, and MHD model performance metrics \citep{Kumar2025}. These works generally suggest that a lower source surface radius of 1.3–2.0 R$_\odot$ provides better agreement with observations, especially at solar maximum. \citet{Panasenco2020} determined the optimal source surface heights for each HCS crossing during Parker Solar Probe's (PSP) orbit, showing that the preferred source surface height varies with longitude.

Magnetic field models often fail to produce photospheric open field regions matching coronal holes and sufficient open flux that matches the in-situ measurements at the same time \citep{Linker2017}. This discrepancy is known as the open flux problem. Current explanations for the open flux problem include an underestimated polar magnetic field \citep{Linker2017, Riley2019} and ignored interchange reconnection along open-closed field boundaries \citep{Antiochos2011, Pontin&Wyper2015, Lockwood2022, Badman2022, Arge2024}.
When using the PFSS model, lowering the source surface height forces more magnetic loops to open and is therefore commonly used to address the open-flux problem. However, simply reducing the source surface height is physically problematic, as it would make coronal loops unrealistically small, thus introducing artificial polarity reversals and spurious current sheets. To overcome this, multiple attempts have been made to consider non-spherical source surfaces (NSSS). \citet{Schulz1978} first explored this idea by employing isogauss surfaces as the source surface for a dipole field star, but the field computations were too complex to be done with spherical harmonic expansion if applied to real solar magnetograms. \citet{Levine1982} further developed this approach by identifying a source surface that minimizes the tangential magnetic field using least-squares methods. \citet{Riley2006} also discussed the sphericity of the source surface by comparing the PFSS model to MHD simulations. \citet{Kruse2020} generalized the PFSS model using an ellipsoidal source surface, enabling a tractable formulation. However, these pioneering studies have not yet led to a fully practical, purely non-spherical model that is readily applicable to solar coronal magnetic field extrapolation.

\begin{figure}[htbp]
    \centering
    \includegraphics[width=0.8\linewidth]{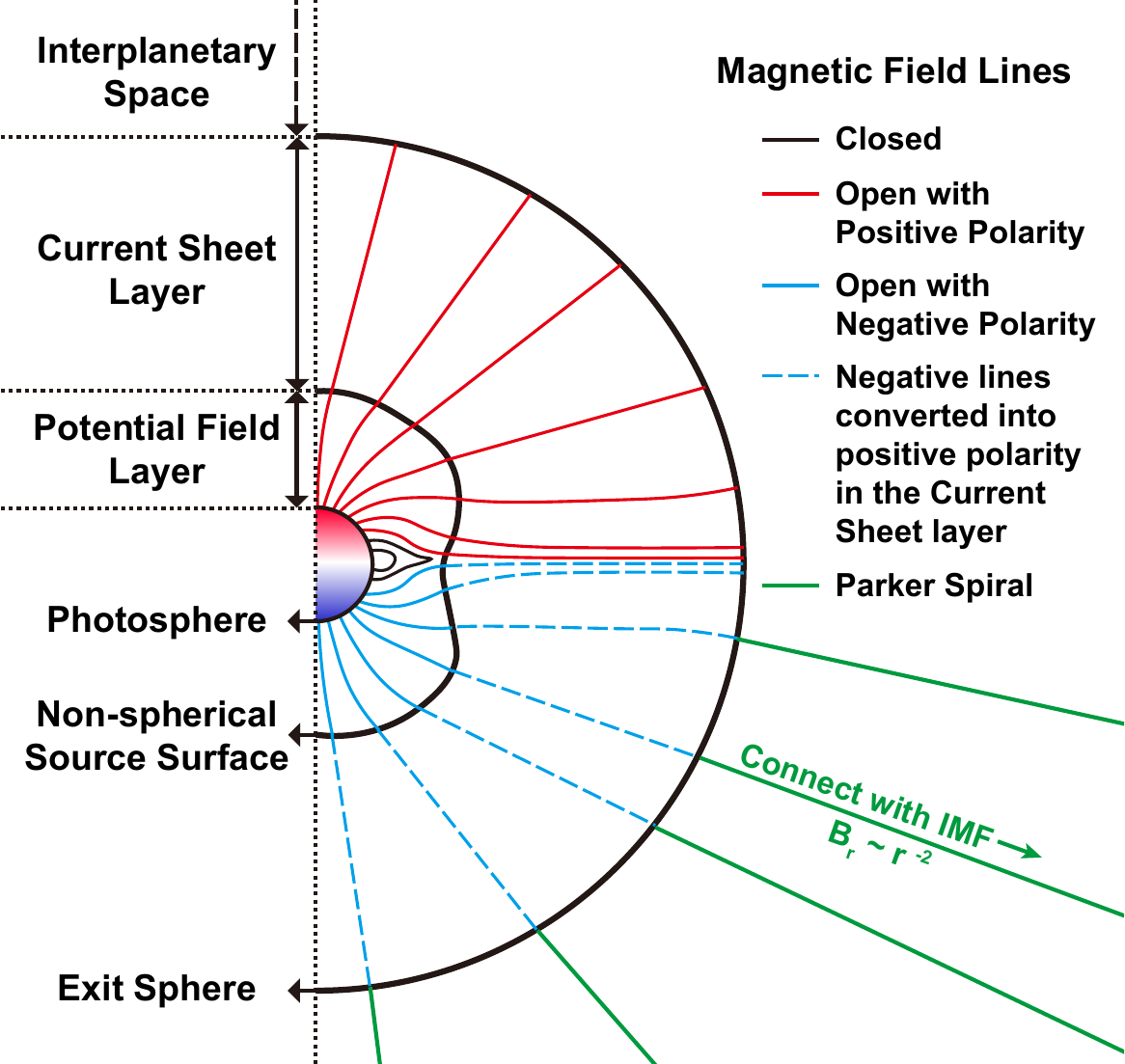}
    \caption{The structure of the NSPF model. The model comprises three concentric layers: (1) a potential field layer; (2) a current sheet layer; (3) an interplanetary layer. The interface between the potential field and current sheet layers is the non-spherical source surface, while the interface between the current sheet and interplanetary layers is the exit sphere (located at 10\(R_\odot\)). Magnetic field lines and layer boundaries are labeled.}
    \label{fig: model_structure}
\end{figure}

In this study, we propose a Non-Spherical Potential Field (NSPF) model by employing the Finite Element Method to solve the Laplace equations with an irregular NSSS. This allows us to achieve more realistic open magnetic flux and coronal loop structures simultaneously. The model structure is illustrated in Fig. \ref{fig: model_structure}, and the model workflow is illustrated in Fig. \ref{fig: workflow} and described in the Method Section. Our model consists of three layers: the potential field layer, the current sheet layer, and the interplanetary layer. The potential field and current sheet layers are separated by the NSSS; the current sheet and interplanetary layers are separated by the exit sphere, located at 10 $R_{\odot}$. The potential field layer uses the synoptic magnetogram as the lower boundary input, and the potential field is calculated within. The current sheet layer takes the potential field layer solution as its lower boundary input, introduces current sheets by first enforcing a uniform positive polarity and then restoring the original polarities after computation, as in the SCS model \citep{Schatten1971}. In the interplanetary layer, magnetic field lines follow the Parker spiral configuration.

We attempt to extract the NSSS as an isosurface of the total magnetic field strength from a magnetic field distribution computed using an initial spherical model. This approach is justified because we expect an ideal NSSS to improve the self-consistency compared with a spherical source surface by capturing two key effects at coronal loop tops: (i) plasma is released when its ram pressure exceeds the magnetic tension that confines it, and (ii) the magnetic potential distribution is modified by the presence of external current sheets. As a result, the NSSS should form concave shapes at the base of helmet streamers, where the magnetic field is weaker. This geometry bends field lines originating from coronal holes toward the current sheets while simultaneously allowing more open field lines near the open–closed field boundary. Consequently, the NSSS approach can produce more realistic open flux budgets and field-line topologies.

We further compare our NSPF model with conventional PFSS and PFSS+PFCS model, and evaluate their performance based on multi-instrument observations. Compared to these models, the NSPF model reproduces complex open-field regions and magnetic topologies that are more consistent with EUV observations. It provides predictions of the IMF magnitude and polarity that match PSP measurements, and yields more compact and physically plausible solar wind source regions simultaneously.

\section{Data and Method}
\subsection{Finite-Element-Method Potential Field Solver}
The Finite-Element-Method Potential Field Solver (FEM-PFS) is the core component of our algorithm, used to solve the Laplace equations in each layer at every iteration. Here we introduce the problem and implementation for FEM-PFS.

\subsubsection{Problem Definition}

Let $\Omega \subset \mathbb{R}^n$ be a domain bounded by a lower boundary $\Gamma_{\rm in}$ and a upper boundary $\Gamma_{\rm out}$, such that $\partial \Omega = \Gamma_{\rm in} \cup \Gamma_{\rm out}$. The Laplace equation for potential field, subject to a Neumann boundary condition on the lower boundary and a Dirichlet boundary condition on the upper boundary, is formulated as:

\begin{equation}
\begin{aligned}
\nabla^2 u &= 0 \quad && \text{in } \Omega \\
u &= 0 \quad && \text{on } \Gamma_{\text{out}} \\
\nabla u \cdot n &= b_n \quad && \text{on } \Gamma_{\text{in}}
\end{aligned}
\end{equation}
where $u$ is the magnetic scalar potential, $b_n$ is the prescribed normal component of the magnetic field on the lower boundary, and $n$ is the outward-pointing unit normal vector perpendicular to $\partial \Omega$.

The equation is computed using DOLFINx \citep{BarattaEtal2023}, the Python extension of the FEniCS Project, which is an open-source package for solving partial differential equations via the finite element method. With DOLFINx, the equation is recast as a variational problem: find $u \in V$ such that

\begin{equation}
    a(u, v)=L(v) \quad \forall v \in V
\end{equation}
where $V$ is an appropriate function space that satisfies the Dirichlet boundary conditions on $\Gamma_{\rm out}$, and the bilinear and linear forms are defined as

\begin{equation}
    \begin{aligned} a(u, v) & :=\int_{\Omega} \nabla u \cdot \nabla v \mathrm{~d} x \\ L(v) & :=\int_{\Gamma_{in}} b_n v \mathrm{~d} s .\end{aligned}    
\end{equation}

With the calculated magnetic scalar potential, the magnetic field is obtained as $\mathbf{B} = \nabla u$, computed with PyVista \citep{sullivan2019pyvista}.

\subsubsection{Mesh Generation and Boundary Conditions}
The potential field layer has the lower boundary as the photosphere and the upper boundary as the source surface. The current sheet layer has the lower boundary as the source surface and the upper boundary as the exit sphere. For each layer, after specifying the lower and upper boundaries, 3D volumetric meshes composed of $\sim$ 800,000 tetrahedra are generated between the boundaries by Gmsh \citep{Gmshpaper}. The surface meshes are refined to the third level, with an approximate angular resolution of $1.8^{\circ}$ and a radial resolution of about $0.03~R_{\odot}$ at $1~R_{\odot}$ and $0.3~R_{\odot}$ at $10~R_{\odot}$. 

For the lower boundary (photosphere) of the potential field layer, the magnetogram is directly interpolated onto each mesh node. For the lower boundary (spherical source surface or NSSS) of the current sheet layer, the magnetic field distribution is reconstructed using 30th-order spherical harmonics and then interpolated onto each mesh node. 

\subsection{Model Routine}
\begin{figure}[htbp]
    \centering
    \includegraphics[width=1.0\linewidth]{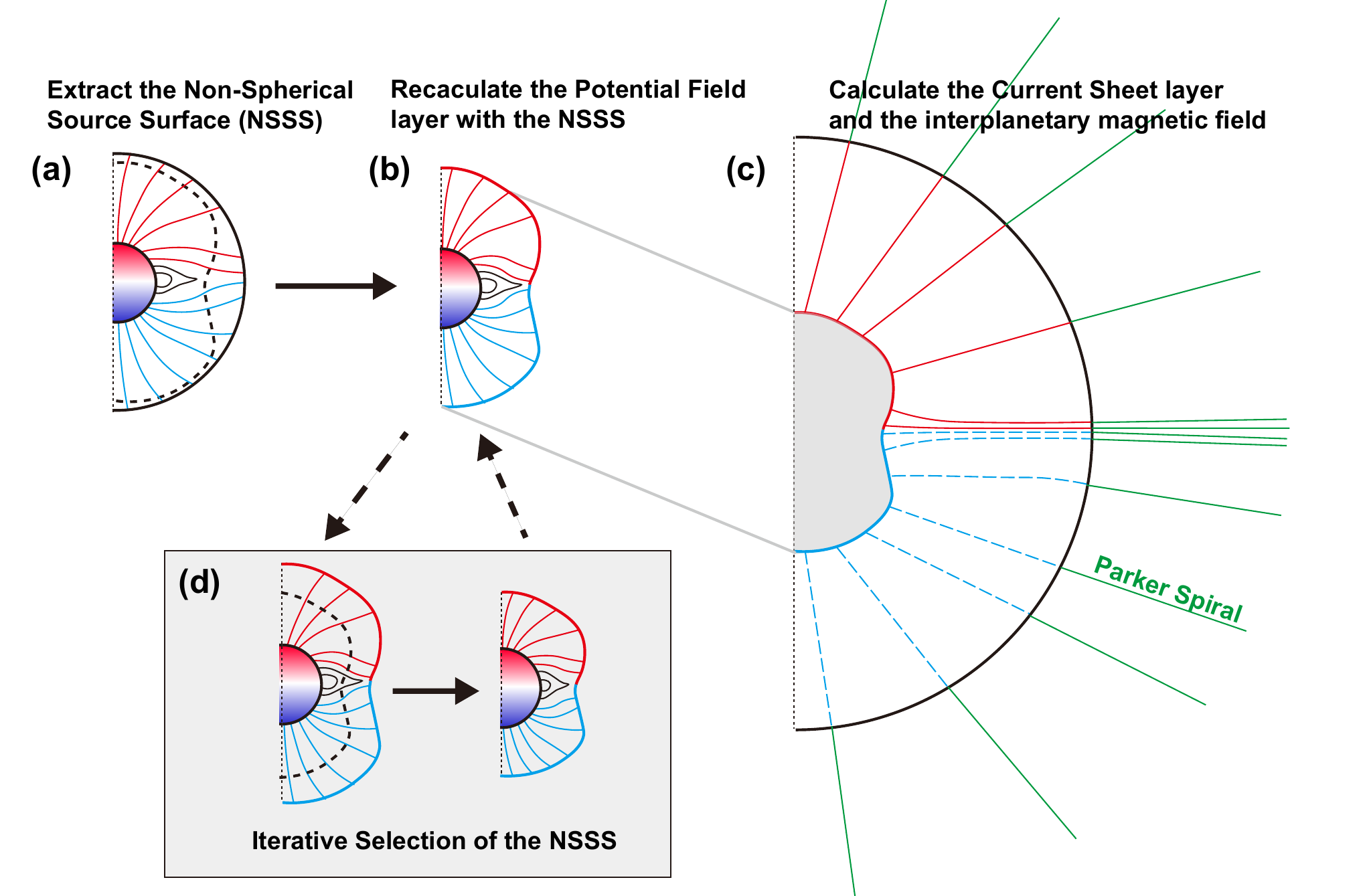}
    \caption{Illustration of the workflow of the NSPF model. (a) extraction of the NSSS from a potential field-layer magnetic field computed with a spherical source surface; (b) recalculation of the potential field-layer magnetic field under the NSSS; (c) computation of the magnetic field in the current sheet layer and the IMF; and (d) optional iteration of the NSSS to match the observed open magnetic flux.}
    \label{fig: workflow}
\end{figure}

In this section, we introduce the workflow to construct the NSPF model from a known magnetogram, as illustrated in Fig. \ref{fig: workflow}. The code is available at \url{https://github.com/wzq215/Non-Spherical_Potential_Field_Model} \citep{Wu2026NSPF}.

\subsubsection{Initial Solution with Spherical Source Surface}
As shown in the Fig. \ref{fig: workflow}a, we first construct a spherical shell with the lower boundary located at the photosphere and the upper boundary at an initial source surface radius, $R_{\mathrm{SS,ini}}$. The normal magnetic field at the lower boundary is prescribed from a measured magnetogram, and the potential field within this layer is calculated using the FEM-PFS. At this stage, the results are equivalent to those obtained from the standard PFSS model with the source surface located at $R_{\mathrm{SS, ini}}$.

\subsubsection{Extraction of the Non-spherical Source Surface}

As shown in the Fig. \ref{fig: workflow}b, within the calculated magnetic field distribution, we extract the NSSS as an isosurface of the magnitude of the magnetic field, $\lvert \mathbf{B}\rvert$. 

By definition, the source surface is supposed to be a surface with zero magnetic potential, where the magnetic field has only normal components on this surface. Thus, all the field lines that penetrate this surface will not return to the Sun, i.e., they are opened up. From the particles’ point of view, ``open up" means that the plasma can escape — becoming a part of the radially expanding solar wind — either when they are not stopped by magnetic pressure (i.e., they are moving along open magnetic flux tubes) or when their ram pressure is higher than the magnetic pressure, usually the magnetic tension of coronal loops that confines them. Above the cusp of coronal loops, the magnetic field is low, and the ratio between plasma thermal pressure and magnetic pressure ($\beta$) is high. Therefore, the plasma can escape and drag the field lines, forming a current sheet. We propose a better NSSS to describe this scenario by forming a concave shape at the base of helmet streamers, where the plasma is easier to escape, and the magnetic potential is distorted by external current sheets. To this end the isosurface of $\lvert \mathbf{B}\rvert$ could serve as a good approximation since it naturally forms a concave shape at the streamer base where the magnetic field is weak, allowing field lines to smoothly converge towards the current sheets as in observations. This concept was first proposed by \citet{Schulz1978} and extended by \citet{Levine1982}. We have also explored another possible isovalue, such as $B_r/\lvert \mathbf{B}\rvert$, which is intended to identify a surface where field lines are quasi-radial. \citet{Riley2006} explored this isosurface in MHD solutions, which would produce a nonspherical surface with average radius around 10~$R_{\odot}$.  However, this choice does not always lead to closed surfaces in a PFSS solution where plasma flows are ignored. By definition, $B_r/\lvert \mathbf{B}\rvert$ equals to 1 at the spherical source surface, but this value drops abruptly near the helmet streamer loops, sometimes leading to source surfaces intersecting the solar surface especially for dipole-like magnetograms.

We first identify the largest possible NSSS within the potential field layer by taking the maximum value of $\lvert \mathbf{B}\rvert$ on the initial spherical source surface as the isovalue for the isosurface. Alternatively, a specific isovalue may be chosen. The isosurface extraction is performed using PyVista. Due to the irregular nature of the unstructured tetrahedral mesh, the interpolation of isosurfaces can introduce geometric artifacts. In particular, small disconnected tetrahedral fragments and highly distorted intersecting triangular facets appear near the boundaries. These artifacts lead to non-physical and non-manifold surface representations, making the extracted isosurface unsuitable for further quantitative analysis. Therefore, we filter out disconnected tetrahedra by only keeping the isosurface with the largest volume. The extracted surface is further smoothed via a 30th-order spherical harmonics reconstruction that preserves the angular resolution of the initial spherical source surface.

\subsubsection{Solution with Non-spherical Source Surface}

We then generate the 3D mesh between the photosphere and the extracted NSSS. The magnetic field inside this domain is recalculated using FEM-PFS. The distribution of the normal magnetic field component, $B_{n}$, on the NSSS is also extracted and reconstructed using 30th-order spherical harmonics for subsequent use.

In this secondary solution, the source surface is defined as the zero-potential surface and therefore is no longer the exact isosurface of the total magnetic field. In principle, one could iterate the procedure to determine a subsequent NSSS. However, we are currently unable to demonstrate that such an iterative approach would converge to a fully self-consistent NSSS—namely, a surface that is simultaneously an isosurface of both the total magnetic field strength and the magnetic potential. In our numerical experiments, the NSSS tends to shrink during the iterative process. The shrinkage occurs because a closed surface can only be extracted from a volume that is smaller than the original computational domain. As a practical alternative, one may constrain the optimal NSSS by requiring that the modeled open magnetic flux match the observed value. Accordingly, in this study, we perform the source-surface extraction only once for each case, varying the initial spherical source-surface radius, and then use observational constraints to select the optimal NSSS configuration. While this approach is not formally rigorous, it is both practical and effective for our purposes.

\subsubsection{Extension to interplanetary space}

Next, we compute the magnetic field distributions within the current sheet layer and the interplanetary space. The 3D mesh of the current sheet layer is generated between the NSSS and the exit sphere. The lower boundary (NSSS) is assigned the normal magnetic field extracted from the magnetic field distribution in the potential field layer, converted to its absolute value. The magnetic field distribution within the current sheet layer is then calculated using FEM-PFS. After the calculation, we restore the correct polarity for each field line in the current sheet layer. The current sheet layer provides a smooth transition between the field lines in the NSPF layer and the Parker spirals in the interplanetary space. Without this layer, field lines extending perpendicularly from the concave regions of the NSSS would intersect with one another. In previous attempts, this issue was avoided by bending the NSSS into a convex shape \citep{Schulz1978,Levine1982}. Here, instead of modifying the NSSS geometry, we resolve this issue by introducing a current sheet layer, which maintains the original NSSS configuration while ensuring a smooth transition to the Parker spiral.

Beyond the exit sphere, magnetic field lines follow Parker spiral, and the magnetic field radial component decreases as $r^{-2}$. The Parker spiral is calculated with the same method as the two-step ballistic backmapping method described in \citet{Bale2019, Hou2024}. The resulted mapping from spacecraft spherical Carrington coordinates $(r_{PSP}, \theta_{PSP}, \phi_{PSP})$ to coordinates on the source surface $(r,\theta,\phi)$ is as follows:
\begin{equation}
\left(\begin{array}{c}
\boldsymbol{r} \\
\theta \\
\phi
\end{array}\right)=\left(\begin{array}{c}
R_{\mathrm{SS}} \\
\theta_{\mathrm{PSP}} \\
\phi_{\mathrm{PSP}}+\frac{\Omega_{\mathrm{S}}}{v_{\mathrm{R}}}\left(r_{\mathrm{PSP}}-R_{\mathrm{ES}}\right)
\end{array}\right)
\end{equation}, where $\Omega_{\mathrm{S}}$ is the solar sidereal rotation rate, $v_R$ is the measured solar wind speed, and $R_{ES}$ is the height of the footpoint on the exit sphere.

\subsection{Data}
We use photospheric magnetograms from the Global Oscillation Network Group (GONG) \citep{Harvey1996} as input for the magnetic field extrapolations.
To provide observational constraints for the model, we incorporate EUV images from the Atmospheric Imaging Assembly (AIA) onboard Solar Dynamics Observatory (SDO) \citep{Lemen2012} and white-light coronagraph images from Large Angle and Spectrometric Coronagraph (LASCO) C2 onboard Solar and Heliospheric Observatory (SOHO) \citep{Brueckner1995}. We also use the SDO/AIA Carrington Map from the dataset provided by \citet{SDO_AIA_CAR_MAP}.
In-situ measurements of the magnetic field from FIELDS \citep{Bale2016}) and solar wind velocity from SWEAP \citep{Kasper2016} onboard the Parker Solar Probe (PSP) \citep{Fox2016} are employed to compare with the extrapolated magnetic field and to perform a two-step ballistic backmapping from the spacecraft to the solar source region.

\section{Results}

We perform coronal magnetic field extrapolation for the Carrington Rotation 2282 (CR 2282; 2024 March 28–31). This period during solar maximum coincides with PSP’s 19th encounter, allowing direct comparison between the model-derived results and in situ observations. Moreover, during this period, PSP is radially aligned with the Earth, making the magnetograms from Earth-based observation more suitable for establishing magnetic connectivity between PSP and the Sun.

As input for the lower boundary, six Global Oscillation Network Group (GONG) magnetograms are selected, each centered at a meridian longitude separated by $10^{\circ}$. For each magnetogram, we run three NSPF models with initial source surface radii $R_{\mathrm{ini}}$ of 2.2 $R_{\odot}$, 2.5 $R_{\odot}$, and 3.0 $R_{\odot}$, and three PFSS models with source surface radii $R_{\mathrm{SS}}$ of 1.5 $R_{\odot}$, 2.0 $R_{\odot}$, and 2.5 $R_{\odot}$. For NSPF models, different NSSSs are extracted from the PFSS solution with different $R_{\mathrm{ini}}$. To compare the extrapolated IMF, we further extend the PFSS solutions into interplanetary space using two approaches: the PFSS + PFCS model and the PFSS + Parker spiral model.
Model performances are evaluated by comparing the predicted field topology with 171~$\AA$ and 193~$\AA$ Extreme Ultraviolet (EUV) images from the Atmospheric Imaging Assembly (AIA) onboard the Solar Dynamics Observatory (SDO) and white-light coronagraph observations from the Large Angle and Spectrometric Coronagraph (LASCO) C2 onboard the Solar and Heliospheric Observatory (SOHO), as well as by comparing the predicted IMF with in-situ magnetic field measurements from PSP.

\subsection{Comparison with EUV and White-light Coronagraphs}

\begin{figure}[!htbp]
    \centering
    \includegraphics[width=1.0\linewidth]{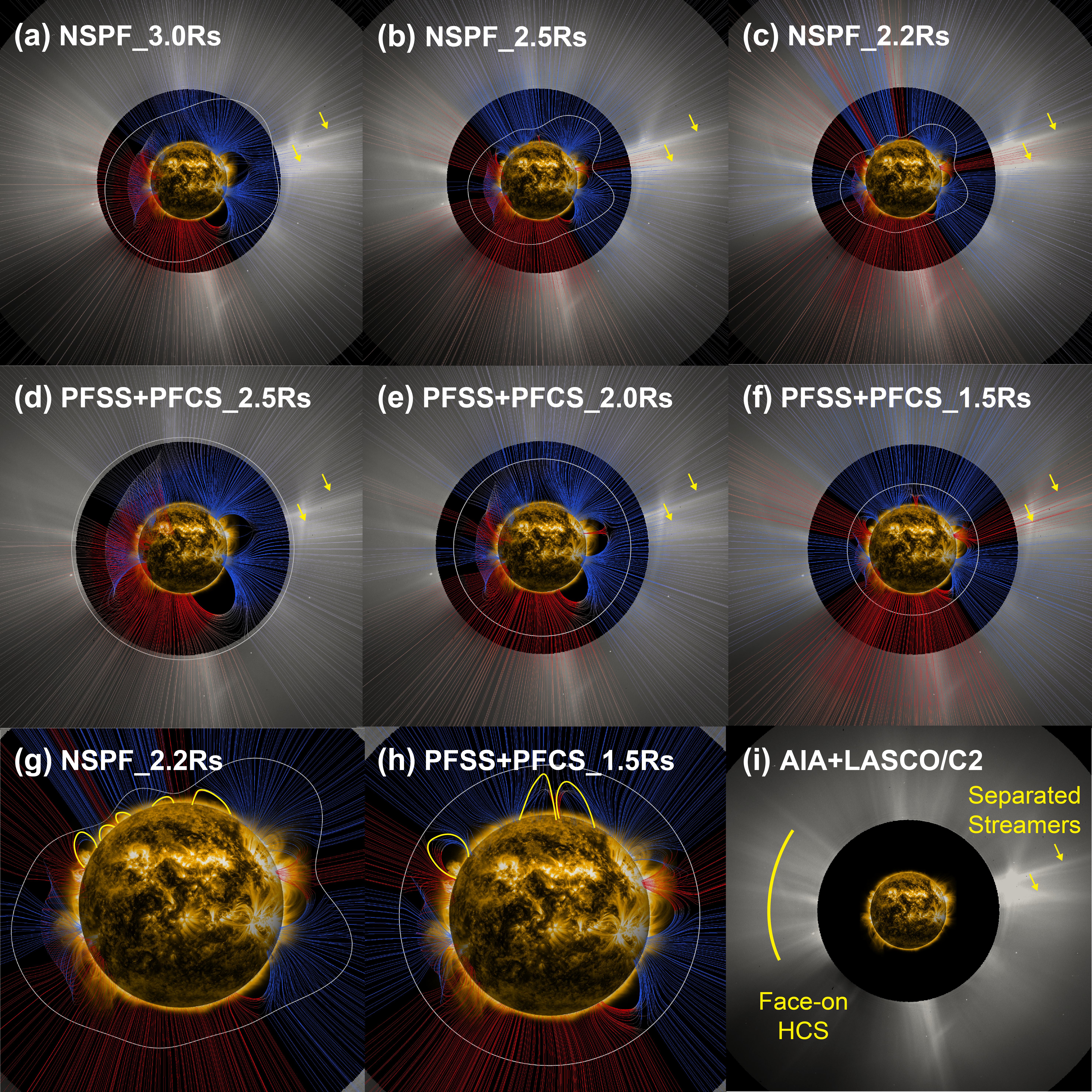}
    \caption{Model results compared with coronagraph observations.
    Panels (a–c) show NSPF models with initial source surface radii of 3.0 $R_{\odot}$, 2.5 $R_{\odot}$, and 2.2 $R_{\odot}$, respectively. Panels (d–f) show PFSS+PFCS models with source surface radii of 2.5 $R_{\odot}$, 2.0 $R_{\odot}$, and 1.5 $R_{\odot}$. Panels (g) and (h) show zoomed-in views of panels (c) and (f), respectively, with coronal loops identified visually and overplotted, allowing a more detailed inspection of the coronal magnetic topology. In each panel, background images are from SDO/AIA 171~$\AA$ and SOHO/LASCO C2, taken on 2024 March 28 at 02:04 UT. The colors of magnetic field lines indicate their polarities. The silver curves represent the boundary of the source surface projected on the image plane. Panel (i) shows the background images, the yellow arc and arrows highlight the face-on HCS and separated streamers, respectively. The yellow arrows in panels (a-f) also refer to the separated streamers.}
    \label{fig: model_vs_euvhi}
\end{figure}

We compare the model-derived magnetic field topology with coronagraph observations, as shown in Fig. \ref{fig: model_vs_euvhi}. For visualization, field lines are traced from seed points distributed along the closed contours defined by the intersections between the sky plane and the surfaces corresponding to 25\%, 50\%, 75\%, and 100\% of the source surface height. A similar approach is adopted in \citet{Benavitz2024} to compare the field lines with eclipse white-light images. In general, among the six models presented in Fig. \ref{fig: model_vs_euvhi}, lower source surface leads to smaller coronal loops and more intricate open-field polarities.

In the LASCO/C2 image, a fan of coronal rays appears on the east limb (yellow arc in Fig. \ref{fig: model_vs_euvhi}i), a bright coronal ray on the south limb, a pair of separated rays on the west limb (yellow arrows in Fig. \ref{fig: model_vs_euvhi}i), and faint fans on the north limb. Based on the modeled topology, the southern ray corresponds to a helmet streamer, while the eastern fan represents streamer belts or HCS viewed face-on.

Different models propose different magnetic structures for the separated rays on the west limb. In the NSPF models with $R_{\mathrm{ini}}=2.5\  R_{\odot}$ and $2.2\ R_{\odot}$, as well as in the PFSS+PFCS model with $R_{\mathrm{SS}}=1.5\ R_{\odot}$ (Figs. \ref{fig: model_vs_euvhi}b, \ref{fig: model_vs_euvhi}c, and \ref{fig: model_vs_euvhi}f), the rays correspond to two distinct helmet streamers originating from source regions with a negative–positive–negative polarity sequence. In contrast, in the other models the rays are associated with a pseudo-streamer configuration. 

This comparison clearly demonstrates that the height of the source surface plays a key role in determining the modeled magnetic structure: models with a lower source surface allow the negative–positive–negative source regions to open, whereas models with a higher source surface prevent magnetic flux from the central positive region from extending outward. Moreover, in NSPF models, the NSSS tends to form a concave shape beneath the base of streamers (see the silver boundary at the west limb in Fig. \ref{fig: model_vs_euvhi}g), allowing more open magnetic flux to extend into interplanetary space and modulating their morphology as well. As a result, the locations of polarity reversals—corresponding to the dense helmet-streamer rays—in the NSPF model with $R_{\mathrm{ini}}=2.2\ R_{\odot}$ provide the best match to the observed coronal rays, even though the true magnetic topology cannot be uniquely diagnosed from white-light observations alone.

A magnified comparison between the EUV images and the modeled magnetic loops in the lower corona for the NSPF model with $R_{\mathrm{ini}}=2.2\ R_{\odot}$ and the PFSS + PFCS model with $R_{\mathrm{SS}}=1.5\ R_{\odot}$ is shown in Figs. \ref{fig: model_vs_euvhi}h and \ref{fig: model_vs_euvhi}g. For the region near the north poles, the NSPF models produce lower loops then PFSS models (overplotted as yellow arcs), resulting open field lines with multiple polarities. The NSPF model produces more complex topologies near the northern pole, where NSSS particularly lowers down and allows multiple small coronal holes to produce open field lines extending into the heliosphere (Fig. \ref{fig: model_vs_euvhi}g).

\subsection{Total Open Magnetic Flux}
We calculate the total open magnetic flux for these models by integrating the magnitude of the field component normal to the source surface,
\begin{equation}
    \Phi_{\mathrm{total}} = \int_{\mathrm{SS}}\lvert B_n\rvert dS .
\end{equation}
To ensure meaningful comparisons across different models, we derive the maximum and minimum source surface radius $R_{\mathrm{SS, max}}$ and $R_{\mathrm{SS, min}}$ for NSSS, as well as the effective source surface radius $R_{\mathrm{SS, eff}}$, defined as the radius of a sphere that encloses the same volume as the NSSS. The results, averaged for six magnetograms in CR 2282, are summarized in Table \ref{tab: open_flux}. As expected, models with a lower $R_{\mathrm{SS, eff}}$ yield higher open flux. The PFSS+PFCS model with $R_{\mathrm{SS}}=1.5\ R_{\odot}$ and the NSPF model with $R_{\mathrm{ini}}=2.2\ R_{\odot}$ produce similar total open flux. However, the NSPF model allows the maximum heights of coronal loops to vary between 0.05 $R_{\odot}$ and 1.08 $R_{\odot}$ above the solar surface across different latitudinal and longitudinal regions. This is more physically realistic than constraining all coronal loops to heights below 0.5 $R_{\odot}$ as in the PFSS + PFCS model. $R_{\mathrm{max}}$ is measured from the reconstructed NSSS rather than prescribed independently. Because the NSSS is extracted using the maximum $|B|$ value on the initial spherical source surface as the reference isovalue, the resulting closed isosurface generally lies inside the initial $R_{\mathrm{SS}}$ in most regions. The spherical-harmonic reconstruction also suppresses small-scale local protrusions of the extracted surface. Therefore, $R_{\mathrm{max}}$ can become smaller than the initial $R_{\mathrm{SS}}$ in the cases considered here.

\begin{table}[!htbp]
\centering
\caption{Comparison of source surface parameters and total open magnetic fluxes between different models.}
\label{tab: open_flux}
\begin{tabular}{lccccc}
\hline
Model & Initial $R_{\mathrm{SS}}$ & $R_{\mathrm{SS,eff}}$ & $R_{\mathrm{SS,max}}$ & $R_{\mathrm{SS,min}}$ & Total open flux \\
      & [$R_{\odot}$]             & [$R_{\odot}$]         & [$R_{\odot}$]         & [$R_{\odot}$]         & [$\mathrm{G\cdot R_{\odot}^2}$] \\
\hline
PFSS + PFCS & 2.5 & 2.5 & 2.5  & 2.5  & 4.64 \\
PFSS + PFCS & 2.0 & 2.0 & 2.0  & 2.0  & 6.96 \\
PFSS + PFCS & 1.5 & 1.5 & 1.5  & 1.5  & 13.65 \\
NSPF      & 3.0 & 2.2 & 2.89 & 1.36 & 6.29 \\
NSPF      & 2.5 & 1.8 & 2.36 & 1.19 & 9.19 \\
NSPF      & 2.2 & 1.6 & 2.08 & 1.05 & 13.01 \\
\hline
\end{tabular}
\\[3pt]
\begin{flushleft}
\textbf{Notes.} 
The initial source surface radius $R_{\mathrm{SS}}$ denotes the radius of the spherical source surface for PFSS+PFCS models, 
and the radius of the initial spherical source surface before extracting the NSSS for NSPF models. 
$R_{\mathrm{SS, eff}}$ is the radius of a sphere enclosing the same volume as the modeled source surface, while 
$R_{\mathrm{SS, max}}$ and $R_{\mathrm{SS, min}}$ are the maximum and minimum radial heliocentric distances of the NSSS, respectively.
\end{flushleft}
\end{table}

\begin{figure}[!htbp]
    \centering
    \includegraphics[width=1.0\linewidth]{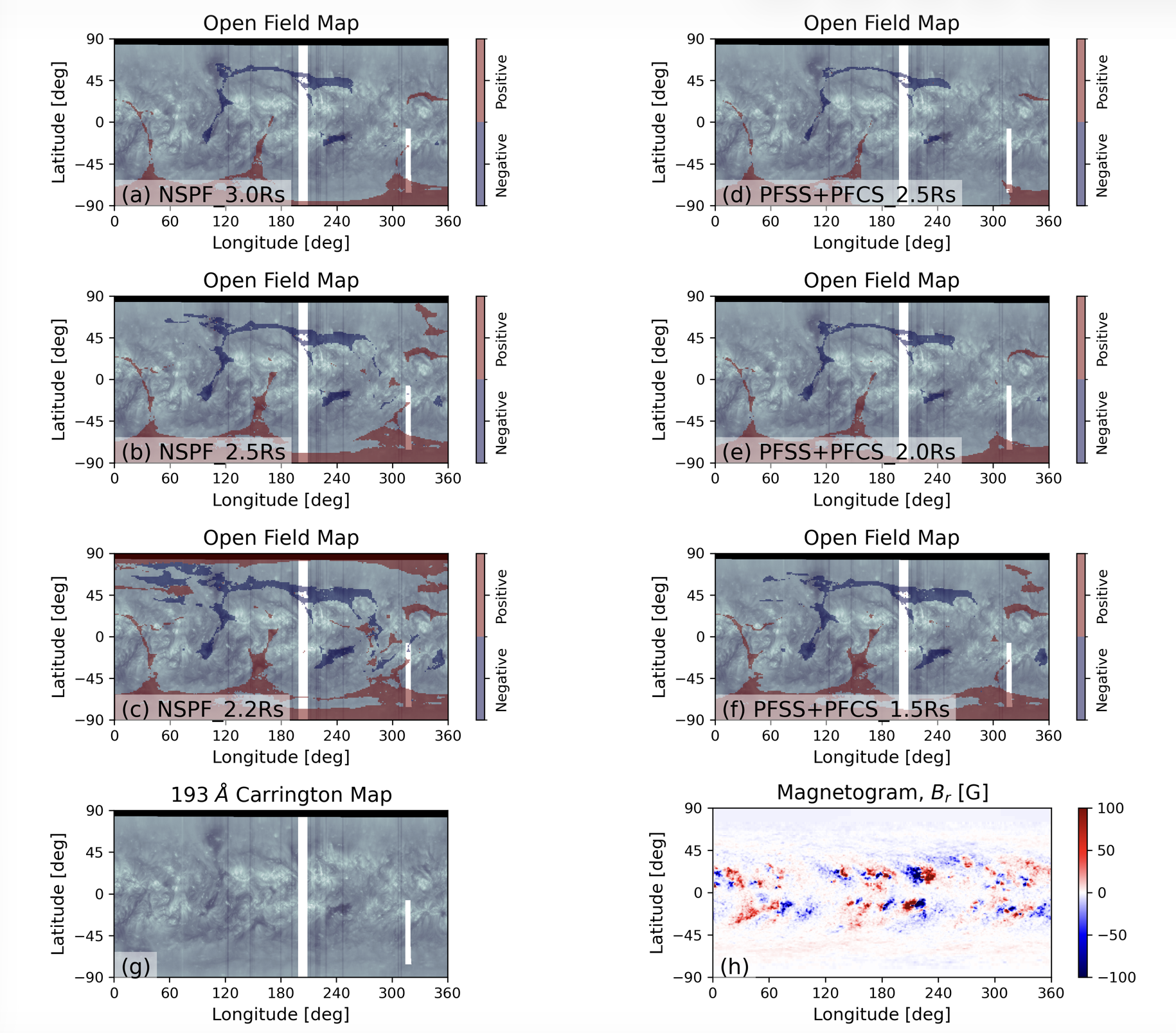}
    \caption{Comparison between the modeled open-field maps and the SDO/AIA Carrington map. The background image is taken from the SDO/AIA 193~$\AA$ Carrington map dataset \citep{SDO_AIA_CAR_MAP}. Red and blue shading indicate positive and negative open-field regions derived from the models, respectively. Panels (a–c) show NSPF models with the initial source surface radii of 3.0~$R_{\odot}$, 2.5~$R_{\odot}$, and 2.2~$R_{\odot}$, while panels (d–f) show PFSS + PFCS models with source surface radii of 2.5~$R_{\odot}$, 2.0~$R_{\odot}$, and 1.5~$R_{\odot}$. Panel (g) shows the bare 193~$\AA$ Carrington map. Panel (h) shows the synoptic magnetogram.}
    \label{fig: open_field_map}
\end{figure}

We further compare the modeled open-field maps with the SDO/AIA 193~$\AA$ Carrington map (Fig. \ref{fig: open_field_map}). The NSPF models produce larger and more complex open-field regions than the PFSS+PFCS models. Moreover, in the NSPF models, the positive and negative open-field regions are closer to each other (e.g., region around $\sim300^{\circ}$ in longitude and $\sim30^{\circ}$ in latitude in Fig. \ref{fig: open_field_map}c). This is because the source surface is located lower above these regions, resulting in smaller closed loops connecting the positive and negative polarities. Consequently, Our leads to more complex polarity inversion lines on the source surface, analogous to that suggested by the Separatrix-Web model \citep{Antiochos2011, Higginson2018}, as smaller streamer-type structures can imprint themselves on the IMF, whereas they would otherwise be enclosed by the global streamers. With more complicated open-field regions, the NSPF models should also yield longer open–closed field boundaries.

The lower source surface in our model may lead to a similar issue as in PFSS models, namely that open-field regions can extend beyond the coronal holes as identified via emission dimming. However, the non-spherical configuration allows for more magnetic flux to be released near streamer bases, supporting the interpretation that interchange reconnection at dynamically evolving open–closed field boundaries can expand the open-field region beyond the extent of coronal holes. Additionally, some artifacts may arise where the source surface is extremely low ($\sim$1.1~$R_{\odot}$) and magnetogram data are less reliable, such as the spurious positive open field near the north pole in Fig.~\ref{fig: open_field_map}(c). The apparent enlargement of polar open-field regions is partly a projection effect.

\subsection{Comparison with In-situ Magnetic Field Measurements}
Our model reproduces global coronal topology consistently with coronagraph observations. We now assess the performance of the compared models by quantifying the extrapolated IMF and comparing it with in-situ spacecraft measurements. 

To model the IMF along a spacecraft’s trajectory from the NSPF models, we employ a two-step ballistic backmapping method. For a given observation time of PSP, a Parker spiral is constructed using the spacecraft’s position and the measured solar wind speed $v_{\mathrm{r, sw}}$. This Parker spiral connects PSP with the exit sphere, where the radial magnetic field $B_{\mathrm{r, ES}}$ is obtained from the model. For simplicity, we use ``$b$'' to denote the radial component of the IMF at PSP's position. The modeled radial component of IMF $b_{\mathrm{NSPF}}$ at PSP’s position is therefore
\begin{equation}
    b_{\mathrm{NSPF}}=B_{r,\mathrm{ES}}\frac{r_{\mathrm{ES}}^2}{r_{\mathrm{PSP}}^2},
\label{Eq_extra}
\end{equation}
where $r_{\mathrm{ES}}$ and $r_{\mathrm{PSP}}$ are the heliocentric distances of field line footpoints on the exit sphere (10 $R_{\odot}$) and PSP, respectively.

For the PFSS + PFCS models, we use the same equation as equation \ref{Eq_extra}, as their only difference from the NSPF models is the assumption of a spherical source surface: 
\begin{equation}
    b_{\mathrm{PFSS+PFCS}}=B_{\mathrm{r, ES}}\frac{r_{\mathrm{ES}}^2}{r_{\mathrm{PSP}}^2}.
\end{equation} 

For the PFSS model without current sheet layers, the modeled IMF is given by 
\begin{equation}
    b_{\mathrm{PFSS}}=B_{\mathrm{r, SS}}\frac{r_{\mathrm{SS}}^2}{r_{\mathrm{PSP}}^2}
\end{equation}
where $r_{\mathrm{SS}}$ is heliocentric distances of source surface footpoints and $B_{\mathrm{r, SS}}$ is the modeled radial magnetic field at source surface footpoints. 
Additionally, for each PSP position, we estimate the solar wind propagation time as $\Delta t = r_{\mathrm{PSP}} / v_{\mathrm{r, sw}}$ and select the magnetogram with the closest timestamp. Here we roughly assume that the solar wind velocity is constant during propagation, which yields a propagation time of $\sim$10 hours. Since the magnetograms are separated by $\sim$18 hours in time or 10$^{\circ}$ in Carrington longitudes, the error in solar wind propagation time is negligible. Crossovers of Parker spirals are inevitable due to the solar wind velocity variation, but in the near-Sun regime, these crossovers have not yet developed forward or backward shocks and therefore they will not be apparently accelerated or decelerated.

\begin{figure}[htbp]
    \centering
    \includegraphics[width=1.0\linewidth]{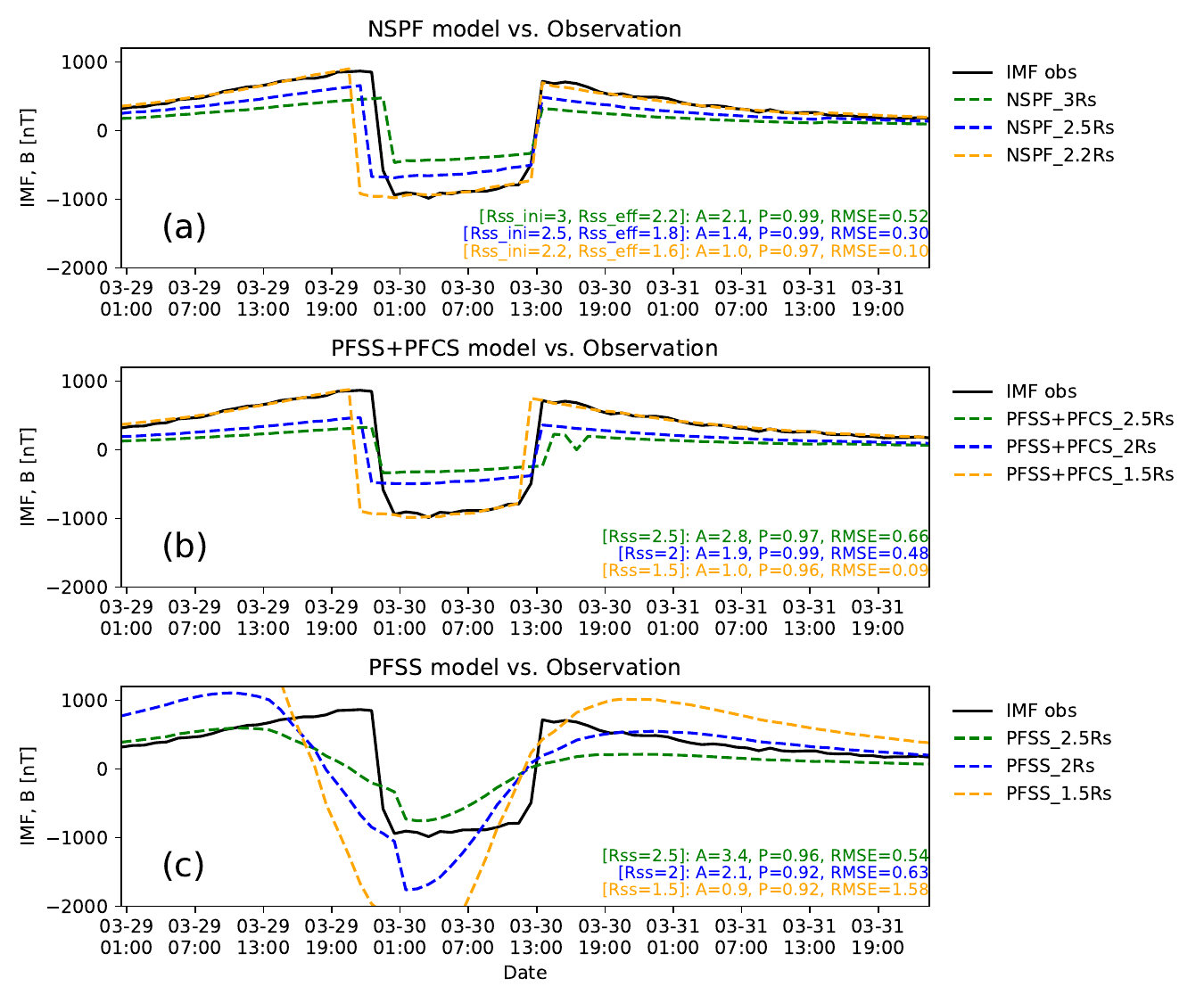}
    \caption{Comparison between the modeled IMF and PSP's in-situ measurements. (a) NSPF models vs. observations. (b) PFSS+PFCS models vs. observations. (c) PFSS models vs. observations. In each panel, the black solid line denotes the observed IMF, while the colored dashed lines represent the modeled IMF obtained from models with different (initial) source surface heights, as indicated in the legend. The text annotations indicate the performance metrics for each model.}
    \label{fig: model_vs_insitu}
\end{figure}

We then compare the modeled radial component of the IMF $b_{\mathrm{model}}$ with the observed radial component of the IMF $b_{\mathrm{obs}}$. We only focus on the radial component since PSP is close to the Sun and Parker spirals are nearly radial. The tangential components are non-zero but not compared in our study since they tend to be more turbulent. The model performance is quantified using three metrics: the average magnitude ratio ($\mathbf{A}$), the correct polarity rate ($\mathbf{P}$), and the relative normalized root mean squared error ($\mathbf{RMSE}$). The definitions of these metrics are provided below, and the corresponding results are presented in Fig. \ref{fig: model_vs_insitu}.

\begin{equation}
    \mathbf A=\frac{1}{N}\sum_{i=1}^{N}\frac{\lvert b_{\mathrm{obs},i}\rvert}{\lvert b_{\mathrm{model},i}\rvert}
\end{equation}

\begin{equation}
    \mathbf P=1-\frac{1}{2N}\sum_{i=1}^{N}\left(1-\frac{b_{\mathrm{obs},i}b_{\mathrm{model},i}}{\lvert b_{\mathrm{obs},i}\rvert \lvert b_{\mathrm{model},i}\rvert}\right)
\end{equation}

\begin{equation}
    \mathbf{RMSE}=\frac{\sqrt{\frac{1}{N}\Sigma_{i=1}^{N}(\lvert b_{\mathrm{obs},i}\rvert r_i^2-\lvert b_{\mathrm{model},i}\rvert r_i^2)^2}}{\frac{1}{N}\Sigma_{i=1}^{N}\lvert b_{\mathrm{obs},i}\rvert r_i^2}    
\end{equation}
where $N$ is the number of sampled time points, $r_i$ represents the heliocentric distance of PSP at the $i$-th sampling point. To incorporate the radial scaling of the magnetic field, we use $\lvert b_{\mathrm{model}, i}\rvert r_i^2$ to calculate the error. These three metrics evaluate the model performance in reproducing the open magnetic flux, field polarity, and magnitude variations, respectively. An ideal model should yield $\mathbf A = 1$, $\mathbf P = 1$, and $\mathbf{RMSE} = 0$. 

Fig. \ref{fig: model_vs_insitu} shows that both the NSPF and PFSS+PFCS models are capable of reproducing the sharp polarity reversals associated with HCS crossings, whereas the PFSS models only yield gradually varying polarities. This difference arises because the current sheet layer in the NSPF and PFSS+PFCS models contains thin current sheets where polarity changes abruptly, as expected in the inner heliosphere. In contrast, in the PFSS models, magnetic field lines are forced to be radial, causing the HCS to diverge beyond the source surface. Although the PFSS+PFCS model with $R_{\mathrm{SS}}=1.5~R_{\odot}$ and the NSPF model with $R_{\mathrm{ini}}=2.2~R_{\odot}$ have similar scores, the NSPF model can better predict the position of polarity reversals, especially for the second HCS crossing around 13:00 on March 30 (Fig. \ref{fig: model_vs_insitu}a). Constrained by PSP's measurements, along with previous comparisons with corona images, we can conclude that the NSPF model with $R_{\mathrm{ini}}=2.2~R_{\odot}$ is the best extrapolation.

\subsection{Source Regions of Solar Wind}
\begin{figure}[htbp]
    \centering
    \includegraphics[width=1.0\linewidth]{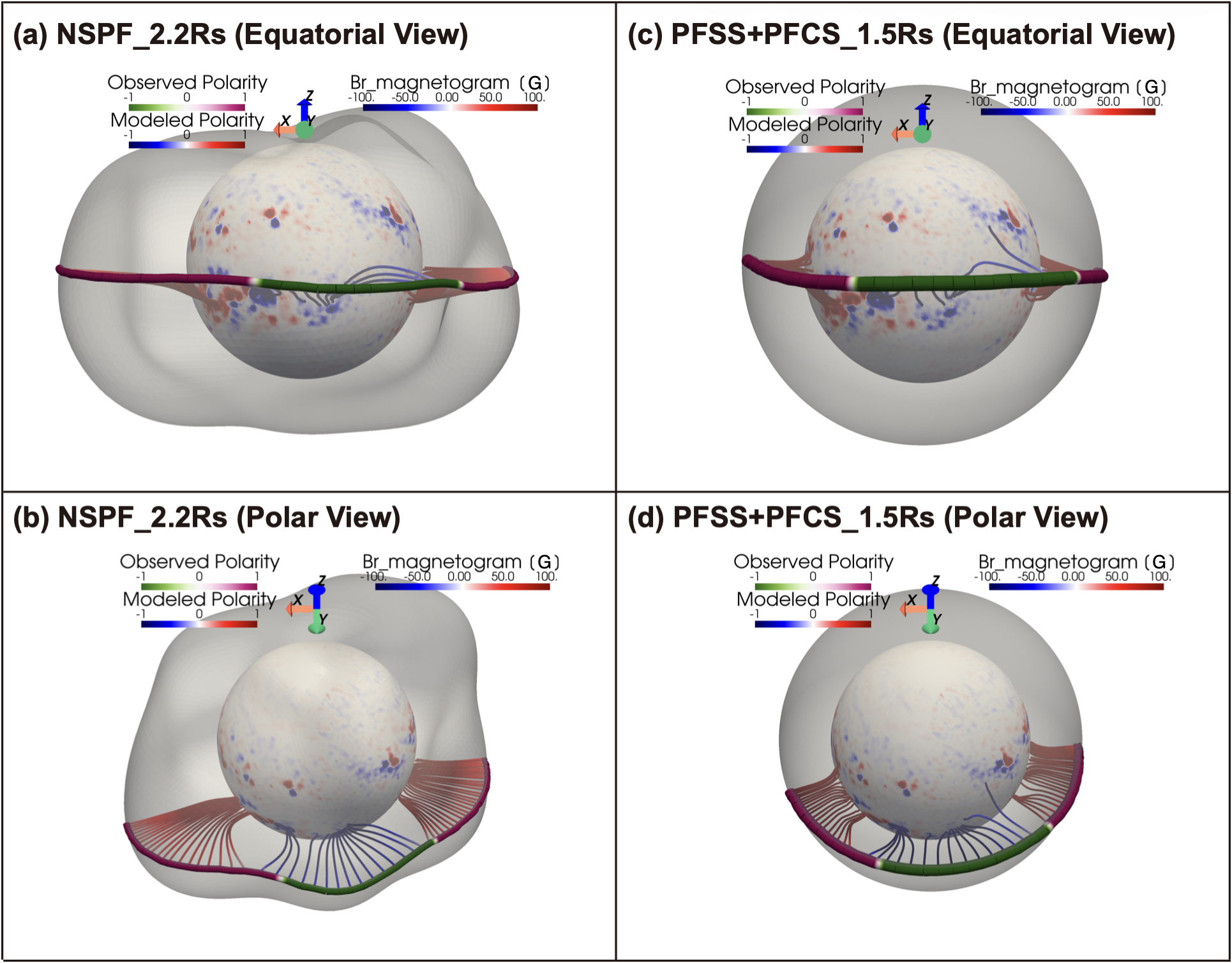}
    \caption{Source regions of the observed solar wind. The inner sphere shows the photospheric radial magnetic field distribution; the outer surface is the source surface. The solid curve on the source surface marks solar wind footpoints, colored by observed magnetic polarity (pink: positive; green: negative). Lines connecting the source surface and the photosphere indicate the modeled magnetic field lines (red: positive; blue: negative). The axes widget indicates the orientation, with the $Z$ axis aligned with the solar rotation axis and the $X$ axis corresponding to zero Carrington longitude. (a), (b): Equatorial and polar view of NSPF model with an initial source surface radius of 2.2 $R_{\odot}$. (c), (d): Equatorial and polar view of PFSS+PFCS model with a source surface radius of 1.5 $R_{\odot}$.  }
    \label{fig: source_region}
\end{figure}

We further investigate the source regions of the solar wind observed by PSP and compare the results between the NSPF model with $R_{\mathrm{ini}}=2.2~R_{\odot}$ and the PFSS+PFCS model with $R_{\mathrm{SS}}=1.5~R_{\odot}$ (Fig. \ref{fig: source_region}). These two models achieve the top two best performances in reproducing the IMF (Fig. \ref{fig: model_vs_insitu}) and exhibit similar total volumes and open magnetic fluxes (Table \ref {tab: open_flux}). 

As shown in Fig. \ref{fig: source_region}, for the solar wind stream with negative polarity, corresponding to that observed by PSP between 01:00 and 13:00 on March 30 (Fig. \ref{fig: model_vs_insitu}), the footpoints produced by the NSPF model are more compact and localized, whereas those from the PFSS+PFCS model tend to spread across different strong-field regions. At times, they extend between mid-latitude coronal holes in the northern and southern hemispheres (see the blue field lines in Fig. \ref{fig: source_region}(c)). This behavior arises because coronal loops beneath streamers in the PFSS+PFCS model are constrained by a prescribed maximum loop height. In the NSPF model, however, the loops are allowed to remain more compact, which is physically more natural and may better match observations, particularly during the solar maximum. Based on the comparison between modeled and observed IMF shown in Fig. \ref{fig: model_vs_insitu}(a, b), the NSPF model can reproduce the observed polarity reversal at 13:00 on March 30 more accurately than the PFSS+PFCS model, suggesting that the source region found by the NSPF model is more realistic.

This indicates that, with a given total open magnetic flux, the NSPF model preferentially opens up field lines near the polarity separatrices where the source surface is lower; while models with a spherical source surface enforce a uniform maximum loop height, which artificially increases the open flux globally and leads to cross-hemisphere drifting of the footpoints. While it is not easy to determine the real source regions of solar wind, our model is more consistent with the scenario that the missing open flux comes from regions near the open-closed field boundaries, possibly dynamically opened through interchange reconnections \citep{Antiochos2011, Pontin&Wyper2015, Badman2022}. 

\section{Discussion and Conclusion}
In this study, we develop a NSPF model as an improvement of the conventional PFSS model. The key idea is to replace the spherical source surface with a non-spherical surface that better represents the geometry of streamer stalks. This approach is motivated by two considerations: (i) allowing a larger fraction of field lines near the cusp of streamer loops to open, thereby releasing more plasma with a high ratio between ram pressure and magnetic pressure, and (ii) self-consistently accounting for modifications to the magnetic potential induced by external current sheets associated with streamers. As a first implementation, we extract an isosurface of the total magnetic field to define a NSSS and solve the corresponding potential field using the finite element method. 

To further account for the heliospheric structure, we wrap the NSPF region with a current sheet layer, following the approach of the SCS model \citep{Schatten1971}, and connect it to interplanetary space with Parker spirals. As a result, with the NSPF model, we establish a comprehensive workflow (Fig. \ref{fig: workflow}) enabling both IMF prediction and solar wind source tracing directly from magnetogram input.

Our NSPF model enhances open magnetic flux locally beneath external current sheets, rather than uniformly lowering the spherical source surface as in PFSS. This geometry preferentially opens flux near the existing open–closed boundary in the PFSS model, especially at the base of helmet streamers. It therefore yields a more physically plausible distribution of the open flux. As a result, for this selected time period, our NSPF model simultaneously achieves (i) a magnetic topology more consistent with coronal observations, (ii) predicted IMF polarities matching in-situ measurements, and (iii) more reliable mapping between solar wind streams and their source regions on the photosphere. While our results are currently evaluated over a limited time interval using PSP data, to address the more general open flux problem, it is required to extend the study to a longer time range, covering different solar activity phases and incorporating in-situ measurements from multiple spacecraft.

Other possible strategies for defining the NSSS include extracting isosurfaces of the angle between the magnetic field and the radial direction, extracting isosurfaces of the magnetic scalar potential of a known magnetic field distribution, or training machine-learning models using solar eclipse observations and MHD simulation outputs. The resulting models could then be further constrained by observations, such as white-light coronagraph images and total open flux derived from IMF observations. In this context, the high-resolution coronagraphic observations between 1.099–3~$R_{\odot}$ provided by the Proba-3 mission offer a unique opportunity to make significant progress on this topic \citep{Zhukov2025}.

Overall, our NSPF model represents a valuable attempt to improve the traditional coronal magnetic field extrapolation models. It can be readily applied to solar wind source region analyses and serves as a proper initial condition for MHD simulations of the solar corona and heliosphere.

\begin{acknowledgements}
The authors acknowledge the contributions of the Parker Solar Probe mission operations and spacecraft engineering teams at the Johns Hopkins University Applied Physics Laboratory and the FIELDS and SWEAP teams for providing the data. We would also like to acknowledge the data provided by GONG and SDO. The authors from China are supported by NSFC, NKRDC, and CNSA under the following grant numbers: 42530105, 42241118, 2021YFA0718600, 2022YFF0503800, 42150105, 42204166, D010301, D050106, and D050103. Z. Wu is also supported by CSC scholarship No. CSC202406010248. This work is also supported by the Specialized Research Fund for the State Key Laboratory of Solar Activity and Space Weather. C. Hou is supported by the Alexander von Humboldt Foundation. D.V. is supported by STFC Consolidated Grant ST/W001004/1. TVD was supported by a Senior Research Project (G088021N) of the FWO Vlaanderen. Furthermore, TVD and Z. Wu received financial support from the Flemish Government under the long-term structural Methusalem funding program, project SOUL: Stellar evolution in full glory, grant METH/24/012 at KU Leuven. The research that led to these results was subsidised by the Belgian Federal Science Policy Office through the contract B2/223/P1/CLOSE-UP. It is also part of the DynaSun project and has thus received funding under the Horizon Europe programme of the European Union under grant agreement (no. 101131534). Views and opinions expressed are however those of the author(s) only and do not necessarily reflect those of the European Union and therefore the European Union cannot be held responsible for them.
\end{acknowledgements}

\bibliography{manuscript}{}

@ARTICLE{Babcock1967,
       author = {{Babcock}, Horace W.},
        title = "{The Zeeman effect in astrophysics}",
      journal = {Physica},
         year = 1967,
        month = jan,
       volume = {33},
       number = {1},
        pages = {102-121},
          doi = {10.1016/0031-8914(67)90263-7},
       adsurl = {https://ui.adsabs.harvard.edu/abs/1967Phy....33..102B}
}

@ARTICLE{Crutcher2019,
       author = {{Crutcher}, Richard M. and {Kemball}, Athol J.},
        title = "{Review of Zeeman Effect Observations of Regions of Star Formation K Zeeman Effect, Magnetic Fields, Star formation, Masers, Molecular clouds}",
      journal = {Frontiers in Astronomy and Space Sciences},
         year = 2019,
        month = oct,
       volume = {6},
          eid = {66},
        pages = {66},
          doi = {10.3389/fspas.2019.00066},
archivePrefix = {arXiv},
       eprint = {1911.06210},
 primaryClass = {astro-ph.GA},
       adsurl = {https://ui.adsabs.harvard.edu/abs/2019FrASS...6...66C}
}

@ARTICLE{Schad2024,
       author = {{Schad}, Thomas A. and {Petrie}, Gordon and {Kuhn}, Jeffrey and {Fehlmann}, Andre and {Rimmele}, Thomas and {Tritschler}, Alexandra and {Woeger}, Friedrich and {Scholl}, Isabelle and {Williams}, Rebecca and {Harrington}, David and {Paraschiv}, Alin and {Szente}, Judit},
        title = "{Mapping the Sun's coronal magnetic field using the Zeeman effect}",
      journal = {Science Advances},
         year = 2024,
        month = sep,
       volume = {10},
       number = {37},
        pages = {eadq1604},
          doi = {10.1126/sciadv.adq1604},
archivePrefix = {arXiv},
       eprint = {2410.21568},
 primaryClass = {astro-ph.SR},
       adsurl = {https://ui.adsabs.harvard.edu/abs/2024SciA...10.1604S}
}

@ARTICLE{Yang2024,
       author = {{Yang}, Zihao and {Tian}, Hui and {Tomczyk}, Steven and {Liu}, Xianyu and {Gibson}, Sarah and {Morton}, Richard J. and {Downs}, Cooper},
        title = "{Observing the evolution of the Sun's global coronal magnetic field over 8 months}",
      journal = {Science},
         year = 2024,
        month = oct,
       volume = {386},
       number = {6717},
        pages = {76-82},
          doi = {10.1126/science.ado2993},
archivePrefix = {arXiv},
       eprint = {2410.16555},
 primaryClass = {astro-ph.SR},
       adsurl = {https://ui.adsabs.harvard.edu/abs/2024Sci...386...76Y}
}

@ARTICLE{Regnier2013,
       author = {{R{\'e}gnier}, S.},
        title = "{Magnetic Field Extrapolations into the Corona: Success and Future Improvements}",
      journal = {\solphys},
         year = 2013,
        month = dec,
       volume = {288},
       number = {2},
        pages = {481-505},
          doi = {10.1007/s11207-013-0367-8},
archivePrefix = {arXiv},
       eprint = {1307.3844},
 primaryClass = {astro-ph.SR},
       adsurl = {https://ui.adsabs.harvard.edu/abs/2013SoPh..288..481R}
}

@ARTICLE{Wiegelmann2017,
       author = {{Wiegelmann}, Thomas and {Petrie}, Gordon J.~D. and {Riley}, Pete},
        title = "{Coronal Magnetic Field Models}",
      journal = {\ssr},
         year = 2017,
        month = sep,
       volume = {210},
       number = {1-4},
        pages = {249-274},
          doi = {10.1007/s11214-015-0178-3},
       adsurl = {https://ui.adsabs.harvard.edu/abs/2017SSRv..210..249W}
}

@ARTICLE{ZhaoHoeksema1994,
       author = {{Zhao}, Xuepu and {Hoeksema}, J. Todd},
        title = "{A Coronal Magnetic Field Model with Horizontal Volume and Sheet Currents}",
      journal = {\solphys},
         year = 1994,
        month = apr,
       volume = {151},
       number = {1},
        pages = {91-105},
          doi = {10.1007/BF00654084},
       adsurl = {https://ui.adsabs.harvard.edu/abs/1994SoPh..151...91Z}
}

@ARTICLE{Mikic2018,
       author = {{Miki{\'c}}, Zoran and {Downs}, Cooper and {Linker}, Jon A. and {Caplan}, Ronald M. and {Mackay}, Duncan H. and {Upton}, Lisa A. and {Riley}, Pete and {Lionello}, Roberto and {T{\"o}r{\"o}k}, Tibor and {Titov}, Viacheslav S. and {Wijaya}, Janvier and {Druckm{\"u}ller}, Miloslav and {Pasachoff}, Jay M. and {Carlos}, Wendy},
        title = "{Predicting the corona for the 21 August 2017 total solar eclipse}",
      journal = {Nature Astronomy},
         year = 2018,
        month = aug,
       volume = {2},
        pages = {913-921},
          doi = {10.1038/s41550-018-0562-5},
       adsurl = {https://ui.adsabs.harvard.edu/abs/2018NatAs...2..913M}
}

@ARTICLE{Wilkins2025,
       author = {{Wilkins}, Chloe P. and {Pontin}, David I. and {Yeates}, Anthony R. and {Antiochos}, Spiro K. and {Schunker}, Hannah and {Lamichhane}, Bishnu},
        title = "{The Sun's Open{\textendash}Closed Flux Boundary and the Origin of the Slow Solar Wind}",
      journal = {\apj},
         year = 2025,
        month = jun,
       volume = {985},
       number = {2},
          eid = {190},
        pages = {190},
          doi = {10.3847/1538-4357/adcd65},
archivePrefix = {arXiv},
       eprint = {2503.09744},
 primaryClass = {astro-ph.SR},
       adsurl = {https://ui.adsabs.harvard.edu/abs/2025ApJ...985..190W}
}

@ARTICLE{Liu2026,
       author = {{Liu}, Xianyu and {Liu}, Weihao and {Manchester}, IV, Ward B. and {Welling}, Daniel T. and {T{\'o}th}, G{\'a}bor and {Gombosi}, Tamas I. and {DeRosa}, Marc L. and {Bertello}, Luca and {Pevtsov}, Alexei A. and {Pevtsov}, Alexander A. and {Reardon}, Kevin and {Wilbanks}, Kathryn and {Rewoldt}, Amy and {Zhao}, Lulu},
        title = "{Simulating the Solar Corona with Multiple Solar Photospheric Magnetic Maps during the 2024 April 8 Total Solar Eclipse}",
      journal = {\apj},
         year = 2026,
        month = feb,
       volume = {997},
       number = {2},
          eid = {243},
        pages = {243},
          doi = {10.3847/1538-4357/ae290f},
       adsurl = {https://ui.adsabs.harvard.edu/abs/2026ApJ...997..243L}
}

@ARTICLE{VDH2014,
       author = {{van der Holst}, B. and {Sokolov}, I.~V. and {Meng}, X. and {Jin}, M. and {Manchester}, IV, W.~B. and {T{\'o}th}, G. and {Gombosi}, T.~I.},
        title = "{Alfv{\'e}n Wave Solar Model (AWSoM): Coronal Heating}",
      journal = {\apj},
         year = 2014,
        month = feb,
       volume = {782},
       number = {2},
          eid = {81},
        pages = {81},
          doi = {10.1088/0004-637X/782/2/81},
archivePrefix = {arXiv},
       eprint = {1311.4093},
 primaryClass = {astro-ph.SR},
       adsurl = {https://ui.adsabs.harvard.edu/abs/2014ApJ...782...81V}
}

@ARTICLE{Wiegelmann2006,
       author = {{Wiegelmann}, T. and {Inhester}, B. and {Kliem}, B. and {Valori}, G. and {Neukirch}, T.},
        title = "{Testing non-linear force-free coronal magnetic field extrapolations with the Titov-D{\'e}moulin equilibrium}",
      journal = {\aap},
         year = 2006,
        month = jul,
       volume = {453},
       number = {2},
        pages = {737-741},
          doi = {10.1051/0004-6361:20054751},
archivePrefix = {arXiv},
       eprint = {astro-ph/0612650},
 primaryClass = {astro-ph},
       adsurl = {https://ui.adsabs.harvard.edu/abs/2006A&A...453..737W}
}

@ARTICLE{Zhu2022,
       author = {{Zhu}, Xiaoshuai and {Neukrich}, Thomas and {Wiegelmann}, Thomas},
        title = "{Magnetohydrostatic modeling of the solar atmosphere}",
      journal = {Science in China E: Technological Sciences},
         year = 2022,
        month = jun,
       volume = {65},
       number = {8},
        pages = {1710-1726},
          doi = {10.1007/s11431-022-2047-8},
archivePrefix = {arXiv},
       eprint = {2203.15356},
 primaryClass = {astro-ph.SR},
       adsurl = {https://ui.adsabs.harvard.edu/abs/2022ScChE..65.1710Z}
}

@ARTICLE{Guo2016A,
       author = {{Guo}, Y. and {Xia}, C. and {Keppens}, R. and {Valori}, G.},
        title = "{Magneto-frictional Modeling of Coronal Nonlinear Force-free Fields. I. Testing with Analytic Solutions}",
      journal = {\apj},
         year = 2016,
        month = sep,
       volume = {828},
       number = {2},
          eid = {82},
        pages = {82},
          doi = {10.3847/0004-637X/828/2/82},
       adsurl = {https://ui.adsabs.harvard.edu/abs/2016ApJ...828...82G}
}

@ARTICLE{Guo2016B,
       author = {{Guo}, Y. and {Xia}, C. and {Keppens}, R.},
        title = "{Magneto-frictional Modeling of Coronal Nonlinear Force-free Fields. II. Application to Observations}",
      journal = {\apj},
         year = 2016,
        month = sep,
       volume = {828},
       number = {2},
          eid = {83},
        pages = {83},
          doi = {10.3847/0004-637X/828/2/83},
       adsurl = {https://ui.adsabs.harvard.edu/abs/2016ApJ...828...83G}
}

@ARTICLE{Wiegelmann&Sakurai2021,
       author = {{Wiegelmann}, Thomas and {Sakurai}, Takashi},
        title = "{Solar force-free magnetic fields}",
      journal = {Living Reviews in Solar Physics},
         year = 2021,
        month = dec,
       volume = {18},
       number = {1},
          eid = {1},
        pages = {1},
          doi = {10.1007/s41116-020-00027-4},
archivePrefix = {arXiv},
       eprint = {1208.4693},
 primaryClass = {astro-ph.SR},
       adsurl = {https://ui.adsabs.harvard.edu/abs/2021LRSP...18....1W}
}

@ARTICLE{Altschuler&Newkirk1969,
       author = {{Altschuler}, Martin D. and {Newkirk}, Jr., Gordon},
        title = "{Magnetic Fields and the Structure of the Solar Corona. I: Methods of Calculating Coronal Fields}",
      journal = {\solphys},
         year = 1969,
        month = sep,
       volume = {9},
       number = {1},
        pages = {131-149},
          doi = {10.1007/BF00145734},
       adsurl = {https://ui.adsabs.harvard.edu/abs/1969SoPh....9..131A}
}

@ARTICLE{Schatten1969,
       author = {{Schatten}, Kenneth H. and {Wilcox}, John M. and {Ness}, Norman F.},
        title = "{A model of interplanetary and coronal magnetic fields}",
      journal = {\solphys},
         year = 1969,
        month = mar,
       volume = {6},
       number = {3},
        pages = {442-455},
          doi = {10.1007/BF00146478},
       adsurl = {https://ui.adsabs.harvard.edu/abs/1969SoPh....6..442S}
}

@ARTICLE{ZhaoHoeksema1995,
       author = {{Zhao}, X. and {Hoeksema}, J.~T.},
        title = "{Predicting the heliospheric magnetic field using the current sheet-source surface model}",
      journal = {Advances in Space Research},
         year = 1995,
        month = aug,
       volume = {16},
       number = {9},
        pages = {181-184},
          doi = {10.1016/0273-1177(95)00331-8},
       adsurl = {https://ui.adsabs.harvard.edu/abs/1995AdSpR..16i.181Z}
}

@BOOK{Aschwanden2005,
       author = {{Aschwanden}, Markus J.},
        title = "{Physics of the Solar Corona. An Introduction with Problems and Solutions (2nd edition)}",
         year = 2005,
          doi = {10.1007/3-540-30766-4},
       adsurl = {https://ui.adsabs.harvard.edu/abs/2005psci.book.....A}
}

@ARTICLE{Stansby2020,
       author = {{Stansby}, David and {Yeates}, Anthony and {Badman}, Samuel},
        title = "{pfsspy: A Python package for potential field source surface modelling}",
      journal = {The Journal of Open Source Software},
         year = 2020,
        month = oct,
       volume = {5},
       number = {54},
          eid = {2732},
        pages = {2732},
          doi = {10.21105/joss.02732},
       adsurl = {https://ui.adsabs.harvard.edu/abs/2020JOSS....5.2732S}
}

@ARTICLE{Gieseler2023,
       author = {{Gieseler}, Jan and {Dresing}, Nina and {Palmroos}, Christian and {Freiherr von Forstner}, Johan L. and {Price}, Daniel J. and {Vainio}, Rami and {Kouloumvakos}, Athanasios and {Rodr{\'\i}guez-Garc{\'\i}a}, Laura and {Trotta}, Domenico and {G{\'e}not}, Vincent and {Masson}, Arnaud and {Roth}, Markus and {Veronig}, Astrid},
        title = "{Solar-MACH: An open-source tool to analyze solar magnetic connection configurations}",
      journal = {Frontiers in Astronomy and Space Sciences},
         year = 2023,
        month = feb,
       volume = {9},
          eid = {384},
        pages = {384},
          doi = {10.3389/fspas.2022.1058810},
archivePrefix = {arXiv},
       eprint = {2210.00819},
 primaryClass = {astro-ph.SR},
       adsurl = {https://ui.adsabs.harvard.edu/abs/2023FrASS...958810G}
}

@ARTICLE{Badman2020,
       author = {{Badman}, Samuel T. and {Bale}, Stuart D. and {Mart{\'\i}nez Oliveros}, Juan C. and {Panasenco}, Olga and {Velli}, Marco and {Stansby}, David and {Buitrago-Casas}, Juan C. and {R{\'e}ville}, Victor and {Bonnell}, John W. and {Case}, Anthony W. and {Dudok de Wit}, Thierry and {Goetz}, Keith and {Harvey}, Peter R. and {Kasper}, Justin C. and {Korreck}, Kelly E. and {Larson}, Davin E. and {Livi}, Roberto and {MacDowall}, Robert J. and {Malaspina}, David M. and {Pulupa}, Marc and {Stevens}, Michael L. and {Whittlesey}, Phyllis L.},
        title = "{Magnetic Connectivity of the Ecliptic Plane within 0.5 au: Potential Field Source Surface Modeling of the First Parker Solar Probe Encounter}",
      journal = {\apjs},
         year = 2020,
        month = feb,
       volume = {246},
       number = {2},
          eid = {23},
        pages = {23},
          doi = {10.3847/1538-4365/ab4da7},
archivePrefix = {arXiv},
       eprint = {1912.02244},
 primaryClass = {astro-ph.SR},
       adsurl = {https://ui.adsabs.harvard.edu/abs/2020ApJS..246...23B}
}

@ARTICLE{Keppens2023,
       author = {{Keppens}, R. and {Popescu Braileanu}, B. and {Zhou}, Y. and {Ruan}, W. and {Xia}, C. and {Guo}, Y. and {Claes}, N. and {Bacchini}, F.},
        title = "{MPI-AMRVAC 3.0: Updates to an open-source simulation framework}",
      journal = {\aap},
         year = 2023,
        month = may,
       volume = {673},
          eid = {A66},
        pages = {A66},
          doi = {10.1051/0004-6361/202245359},
archivePrefix = {arXiv},
       eprint = {2303.03026},
 primaryClass = {astro-ph.IM},
       adsurl = {https://ui.adsabs.harvard.edu/abs/2023A&A...673A..66K}
}

@INPROCEEDINGS{Arge2003,
       author = {{Arge}, Charles N. and {Odstrcil}, Dusan and {Pizzo}, Victor J. and {Mayer}, Leslie R.},
        title = "{Improved Method for Specifying Solar Wind Speed Near the Sun}",
    booktitle = {Solar Wind Ten},
         year = 2003,
       editor = {{Velli}, Marco and {Bruno}, Roberto and {Malara}, Francesco and {Bucci}, B.},
       series = {American Institute of Physics Conference Series},
       volume = {679},
        month = sep,
    publisher = {AIP},
        pages = {190-193},
          doi = {10.1063/1.1618574},
       adsurl = {https://ui.adsabs.harvard.edu/abs/2003AIPC..679..190A}
}

@ARTICLE{Knizhnik2024,
       author = {{Knizhnik}, Kalman J.},
        title = "{The Schatten current sheet}",
      journal = {Frontiers in Astronomy and Space Sciences},
         year = 2024,
        month = oct,
       volume = {11},
          eid = {1476498},
        pages = {1476498},
          doi = {10.3389/fspas.2024.1476498},
       adsurl = {https://ui.adsabs.harvard.edu/abs/2024FrASS..1176498K}
}

@ARTICLE{Schatten1971,
       author = {{Schatten}, K.~H.},
        title = "{Current sheet magnetic model for the solar corona.}",
      journal = {Cosmic Electrodynamics},
         year = 1971,
        month = jan,
       volume = {2},
        pages = {232-245},
       adsurl = {https://ui.adsabs.harvard.edu/abs/1971CosEl...2..232S}
}

@ARTICLE{Koskela2019,
       author = {{Koskela}, Jennimari and {Virtanen}, Ilpo and {Mursula}, Kalevi},
        title = "{Revisiting the coronal current sheet model: Parameter range analysis and comparison with the potential field model}",
      journal = {\aap},
         year = 2019,
        month = nov,
       volume = {631},
          eid = {A17},
        pages = {A17},
          doi = {10.1051/0004-6361/201935967},
       adsurl = {https://ui.adsabs.harvard.edu/abs/2019A&A...631A..17K}
}

@ARTICLE{Benavitz2024,
       author = {{Benavitz}, Luke Fushimi and {Boe}, Benjamin and {Habbal}, Shadia Rifai},
        title = "{Total Solar Eclipse White-light Images as a Benchmark for Potential Field Source Surface Coronal Magnetic Field Models: An In-depth Analysis over a Solar Cycle}",
      journal = {\apj},
         year = 2024,
        month = oct,
       volume = {974},
       number = {2},
          eid = {178},
        pages = {178},
          doi = {10.3847/1538-4357/ad71c6},
archivePrefix = {arXiv},
       eprint = {2408.16149},
 primaryClass = {astro-ph.SR},
       adsurl = {https://ui.adsabs.harvard.edu/abs/2024ApJ...974..178B}
}

@ARTICLE{Shi2024,
       author = {{Shi}, Guanglu and {Feng}, Li and {Ying}, Beili and {Li}, Shuting and {Gan}, Weiqun},
        title = "{Refinement of Global Coronal and Interplanetary Magnetic Field Extrapolations Constrained by Remote-sensing and In Situ Observations at the Solar Minimum}",
      journal = {\apj},
         year = 2024,
        month = aug,
       volume = {970},
       number = {2},
          eid = {131},
        pages = {131},
          doi = {10.3847/1538-4357/ad5200},
archivePrefix = {arXiv},
       eprint = {2405.18665},
 primaryClass = {astro-ph.SR},
       adsurl = {https://ui.adsabs.harvard.edu/abs/2024ApJ...970..131S}
}

@ARTICLE{Arden2014,
       author = {{Arden}, W.~M. and {Norton}, A.~A. and {Sun}, X.},
        title = "{A ``breathing'' source surface for cycles 23 and 24}",
      journal = {Journal of Geophysical Research (Space Physics)},
         year = 2014,
        month = mar,
       volume = {119},
       number = {3},
        pages = {1476-1485},
          doi = {10.1002/2013JA019464},
       adsurl = {https://ui.adsabs.harvard.edu/abs/2014JGRA..119.1476A}
}

@INPROCEEDINGS{Lee2010,
       author = {{Lee}, Christina O. and {Luhmann}, Janet G. and {Hoeksema}, J. Todd and {Sun}, Xudong and {de Pater}, Imke},
        title = "{Coronal Field Opens at Lower Height During the Weak Solar Cycle 23 Minimum: IMF Comparison Suggests Source Surface Should be Lowered to 1.8 Solar Radii}",
    booktitle = {Solar Heliospheric and INterplanetary Environment (SHINE 2010)},
         year = 2010,
        month = jul,
          eid = {10},
        pages = {10},
       adsurl = {https://ui.adsabs.harvard.edu/abs/2010shin.confE..10L}
}

@ARTICLE{Shoda2025,
       author = {{Shoda}, Munehito and {Tokoro}, Kyogo and {Shiota}, Daikou and {Imada}, Shinsuke},
        title = "{Empirical Optimization of the Source-surface Height in the Potential Field Source Surface Extrapolation}",
      journal = {\apj},
         year = 2025,
        month = nov,
       volume = {993},
       number = {2},
          eid = {242},
        pages = {242},
          doi = {10.3847/1538-4357/ae10ba},
archivePrefix = {arXiv},
       eprint = {2510.05513},
 primaryClass = {astro-ph.SR},
       adsurl = {https://ui.adsabs.harvard.edu/abs/2025ApJ...993..242S}
}

@ARTICLE{Kumar2025,
       author = {{Kumar}, Sandeep and {Srivastava}, Nandita and {Talpeanu}, Dana-Camelia and {Mierla}, Marilena and {D'Huys}, Elke and {Dominique}, Marie},
        title = "{On the role of source surface height and magnetograms in solar wind forecast accuracy}",
      journal = {Journal of Space Weather and Space Climate},
         year = 2025,
        month = jan,
       volume = {15},
          eid = {24},
        pages = {24},
          doi = {10.1051/swsc/2025021},
       adsurl = {https://ui.adsabs.harvard.edu/abs/2025JSWSC..15...24K}
}

@ARTICLE{Badman2022,
       author = {{Badman}, Samuel T. and {Brooks}, David H. and {Poirier}, Nicolas and {Warren}, Harry P. and {Petrie}, Gordon and {Rouillard}, Alexis P. and {Nick Arge}, C. and {Bale}, Stuart D. and {de Pablos Ag{\"u}ero}, Diego and {Harra}, Louise and {Jones}, Shaela I. and {Kouloumvakos}, Athanasios and {Riley}, Pete and {Panasenco}, Olga and {Velli}, Marco and {Wallace}, Samantha},
        title = "{Constraining Global Coronal Models with Multiple Independent Observables}",
      journal = {\apj},
         year = 2022,
        month = jun,
       volume = {932},
       number = {2},
          eid = {135},
        pages = {135},
          doi = {10.3847/1538-4357/ac6610},
archivePrefix = {arXiv},
       eprint = {2201.11818},
 primaryClass = {astro-ph.SR},
       adsurl = {https://ui.adsabs.harvard.edu/abs/2022ApJ...932..135B}
}

@ARTICLE{Schulz1978,
       author = {{Schulz}, M. and {Frazier}, E.~N. and {Boucher}, Jr., D.~J.},
        title = "{Coronal magnetic-field model with non-spherical source surface.}",
      journal = {\solphys},
         year = 1978,
        month = nov,
       volume = {60},
       number = {1},
        pages = {83-104},
          doi = {10.1007/BF00152334},
       adsurl = {https://ui.adsabs.harvard.edu/abs/1978SoPh...60...83S}
}

@ARTICLE{Kruse2020,
       author = {{Kruse}, M. and {Heidrich-Meisner}, V. and {Wimmer-Schweingruber}, R.~F. and {Hauptmann}, M.},
        title = "{An elliptic expansion of the potential field source surface model}",
      journal = {\aap},
         year = 2020,
        month = jun,
       volume = {638},
          eid = {A109},
        pages = {A109},
          doi = {10.1051/0004-6361/202037734},
archivePrefix = {arXiv},
       eprint = {2005.12843},
 primaryClass = {astro-ph.SR},
       adsurl = {https://ui.adsabs.harvard.edu/abs/2020A&A...638A.109K}
}

@ARTICLE{Harvey1996,
       author = {{Harvey}, J.~W. and {Hill}, F. and {Hubbard}, R.~P. and {Kennedy}, J.~R. and {Leibacher}, J.~W. and {Pintar}, J.~A. and {Gilman}, P.~A. and {Noyes}, R.~W. and {Title}, A.~M. and {Toomre}, J. and {Ulrich}, R.~K. and {Bhatnagar}, A. and {Kennewell}, J.~A. and {Marquette}, W. and {Patron}, J. and {Saa}, O. and {Yasukawa}, E.},
        title = "{The Global Oscillation Network Group (GONG) Project}",
      journal = {Science},
         year = 1996,
        month = may,
       volume = {272},
       number = {5266},
        pages = {1284-1286},
          doi = {10.1126/science.272.5266.1284},
       adsurl = {https://ui.adsabs.harvard.edu/abs/1996Sci...272.1284H}
}

@ARTICLE{Lemen2012,
       author = {{Lemen}, James R. and {Title}, Alan M. and {Akin}, David J. and {Boerner}, Paul F. and {Chou}, Catherine and {Drake}, Jerry F. and {Duncan}, Dexter W. and {Edwards}, Christopher G. and {Friedlaender}, Frank M. and {Heyman}, Gary F. and {Hurlburt}, Neal E. and {Katz}, Noah L. and {Kushner}, Gary D. and {Levay}, Michael and {Lindgren}, Russell W. and {Mathur}, Dnyanesh P. and {McFeaters}, Edward L. and {Mitchell}, Sarah and {Rehse}, Roger A. and {Schrijver}, Carolus J. and {Springer}, Larry A. and {Stern}, Robert A. and {Tarbell}, Theodore D. and {Wuelser}, Jean-Pierre and {Wolfson}, C. Jacob and {Yanari}, Carl and {Bookbinder}, Jay A. and {Cheimets}, Peter N. and {Caldwell}, David and {Deluca}, Edward E. and {Gates}, Richard and {Golub}, Leon and {Park}, Sang and {Podgorski}, William A. and {Bush}, Rock I. and {Scherrer}, Philip H. and {Gummin}, Mark A. and {Smith}, Peter and {Auker}, Gary and {Jerram}, Paul and {Pool}, Peter and {Soufli}, Regina and {Windt}, David L. and {Beardsley}, Sarah and {Clapp}, Matthew and {Lang}, James and {Waltham}, Nicholas},
        title = "{The Atmospheric Imaging Assembly (AIA) on the Solar Dynamics Observatory (SDO)}",
      journal = {\solphys},
         year = 2012,
        month = jan,
       volume = {275},
       number = {1-2},
        pages = {17-40},
          doi = {10.1007/s11207-011-9776-8},
       adsurl = {https://ui.adsabs.harvard.edu/abs/2012SoPh..275...17L}
}

@ARTICLE{Brueckner1995,
       author = {{Brueckner}, G.~E. and {Howard}, R.~A. and {Koomen}, M.~J. and {Korendyke}, C.~M. and {Michels}, D.~J. and {Moses}, J.~D. and {Socker}, D.~G. and {Dere}, K.~P. and {Lamy}, P.~L. and {Llebaria}, A. and {Bout}, M.~V. and {Schwenn}, R. and {Simnett}, G.~M. and {Bedford}, D.~K. and {Eyles}, C.~J.},
        title = "{The Large Angle Spectroscopic Coronagraph (LASCO)}",
      journal = {\solphys},
         year = 1995,
        month = dec,
       volume = {162},
       number = {1-2},
        pages = {357-402},
          doi = {10.1007/BF00733434},
       adsurl = {https://ui.adsabs.harvard.edu/abs/1995SoPh..162..357B}
}

@ARTICLE{Fox2016,
       author = {{Fox}, N.~J. and {Velli}, M.~C. and {Bale}, S.~D. and {Decker}, R. and {Driesman}, A. and {Howard}, R.~A. and {Kasper}, J.~C. and {Kinnison}, J. and {Kusterer}, M. and {Lario}, D. and {Lockwood}, M.~K. and {McComas}, D.~J. and {Raouafi}, N.~E. and {Szabo}, A.},
        title = "{The Solar Probe Plus Mission: Humanity's First Visit to Our Star}",
      journal = {\ssr},
         year = 2016,
        month = dec,
       volume = {204},
       number = {1-4},
        pages = {7-48},
          doi = {10.1007/s11214-015-0211-6},
       adsurl = {https://ui.adsabs.harvard.edu/abs/2016SSRv..204....7F}
}

@ARTICLE{Bale2016,
       author = {{Bale}, S.~D. and {Goetz}, K. and {Harvey}, P.~R. and {Turin}, P. and {Bonnell}, J.~W. and {Dudok de Wit}, T. and {Ergun}, R.~E. and {MacDowall}, R.~J. and {Pulupa}, M. and {Andre}, M. and {Bolton}, M. and {Bougeret}, J.-L. and {Bowen}, T.~A. and {Burgess}, D. and {Cattell}, C.~A. and {Chandran}, B.~D.~G. and {Chaston}, C.~C. and {Chen}, C.~H.~K. and {Choi}, M.~K. and {Connerney}, J.~E. and {Cranmer}, S. and {Diaz-Aguado}, M. and {Donakowski}, W. and {Drake}, J.~F. and {Farrell}, W.~M. and {Fergeau}, P. and {Fermin}, J. and {Fischer}, J. and {Fox}, N. and {Glaser}, D. and {Goldstein}, M. and {Gordon}, D. and {Hanson}, E. and {Harris}, S.~E. and {Hayes}, L.~M. and {Hinze}, J.~J. and {Hollweg}, J.~V. and {Horbury}, T.~S. and {Howard}, R.~A. and {Hoxie}, V. and {Jannet}, G. and {Karlsson}, M. and {Kasper}, J.~C. and {Kellogg}, P.~J. and {Kien}, M. and {Klimchuk}, J.~A. and {Krasnoselskikh}, V.~V. and {Krucker}, S. and {Lynch}, J.~J. and {Maksimovic}, M. and {Malaspina}, D.~M. and {Marker}, S. and {Martin}, P. and {Martinez-Oliveros}, J. and {McCauley}, J. and {McComas}, D.~J. and {McDonald}, T. and {Meyer-Vernet}, N. and {Moncuquet}, M. and {Monson}, S.~J. and {Mozer}, F.~S. and {Murphy}, S.~D. and {Odom}, J. and {Oliverson}, R. and {Olson}, J. and {Parker}, E.~N. and {Pankow}, D. and {Phan}, T. and {Quataert}, E. and {Quinn}, T. and {Ruplin}, S.~W. and {Salem}, C. and {Seitz}, D. and {Sheppard}, D.~A. and {Siy}, A. and {Stevens}, K. and {Summers}, D. and {Szabo}, A. and {Timofeeva}, M. and {Vaivads}, A. and {Velli}, M. and {Yehle}, A. and {Werthimer}, D. and {Wygant}, J.~R.},
        title = "{The FIELDS Instrument Suite for Solar Probe Plus. Measuring the Coronal Plasma and Magnetic Field, Plasma Waves and Turbulence, and Radio Signatures of Solar Transients}",
      journal = {\ssr},
         year = 2016,
        month = dec,
       volume = {204},
       number = {1-4},
        pages = {49-82},
          doi = {10.1007/s11214-016-0244-5},
       adsurl = {https://ui.adsabs.harvard.edu/abs/2016SSRv..204...49B}
}

@ARTICLE{Kasper2016,
       author = {{Kasper}, Justin C. and {Abiad}, Robert and {Austin}, Gerry and {Balat-Pichelin}, Marianne and {Bale}, Stuart D. and {Belcher}, John W. and {Berg}, Peter and {Bergner}, Henry and {Berthomier}, Matthieu and {Bookbinder}, Jay and {Brodu}, Etienne and {Caldwell}, David and {Case}, Anthony W. and {Chandran}, Benjamin D.~G. and {Cheimets}, Peter and {Cirtain}, Jonathan W. and {Cranmer}, Steven R. and {Curtis}, David W. and {Daigneau}, Peter and {Dalton}, Greg and {Dasgupta}, Brahmananda and {DeTomaso}, David and {Diaz-Aguado}, Millan and {Djordjevic}, Blagoje and {Donaskowski}, Bill and {Effinger}, Michael and {Florinski}, Vladimir and {Fox}, Nichola and {Freeman}, Mark and {Gallagher}, Dennis and {Gary}, S. Peter and {Gauron}, Tom and {Gates}, Richard and {Goldstein}, Melvin and {Golub}, Leon and {Gordon}, Dorothy A. and {Gurnee}, Reid and {Guth}, Giora and {Halekas}, Jasper and {Hatch}, Ken and {Heerikuisen}, Jacob and {Ho}, George and {Hu}, Qiang and {Johnson}, Greg and {Jordan}, Steven P. and {Korreck}, Kelly E. and {Larson}, Davin and {Lazarus}, Alan J. and {Li}, Gang and {Livi}, Roberto and {Ludlam}, Michael and {Maksimovic}, Milan and {McFadden}, James P. and {Marchant}, William and {Maruca}, Bennet A. and {McComas}, David J. and {Messina}, Luciana and {Mercer}, Tony and {Park}, Sang and {Peddie}, Andrew M. and {Pogorelov}, Nikolai and {Reinhart}, Matthew J. and {Richardson}, John D. and {Robinson}, Miles and {Rosen}, Irene and {Skoug}, Ruth M. and {Slagle}, Amanda and {Steinberg}, John T. and {Stevens}, Michael L. and {Szabo}, Adam and {Taylor}, Ellen R. and {Tiu}, Chris and {Turin}, Paul and {Velli}, Marco and {Webb}, Gary and {Whittlesey}, Phyllis and {Wright}, Ken and {Wu}, S.~T. and {Zank}, Gary},
        title = "{Solar Wind Electrons Alphas and Protons (SWEAP) Investigation: Design of the Solar Wind and Coronal Plasma Instrument Suite for Solar Probe Plus}",
      journal = {\ssr},
         year = 2016,
        month = dec,
       volume = {204},
       number = {1-4},
        pages = {131-186},
          doi = {10.1007/s11214-015-0206-3},
       adsurl = {https://ui.adsabs.harvard.edu/abs/2016SSRv..204..131K}
}

@misc{SDO_AIA_CAR_MAP, title={SDO/AIA Carrington Maps}, url={https://spase-metadata.org/NASA/DisplayData/SDO/AIA/SSC/CarringtonMaps.html}, DOI={10.48322/205Z-N617}, abstractNote={Synoptic (or Carrington) Map PNG images created from EUV images observed by the Atmospheric Imaging Assembly (AIA) on the Solar Dynamics Observatory (SDO). Maps are available for each Carrington Rotation since the beginning of the SDO mission. Meridian maps for AIA illustrate the corresponding structures and long-term changes in the lower atmosphere.}, publisher={STEREO Science Center}, author={Thernisien, Arnaud and Hutting, Lynn and Ugarte-Urra, Ignacio and Rich, Nathan B. and Howard, Russell A. and Wood, Brian}, year={2024} , version={Accessed on 2025-10-27}}

@ARTICLE{Hou2024,
       author = {{Hou}, Chuanpeng and {He}, Jiansen and {Duan}, Die and {Wu}, Ziqi and {Chen}, Yajie and {Verscharen}, Daniel and {Rouillard}, Alexis P. and {Li}, Huichao and {Yang}, Liping and {Bale}, Stuart D.},
        title = "{The origin of interplanetary switchbacks in reconnection at chromospheric network boundaries}",
      journal = {Nature Astronomy},
         year = 2024,
        month = oct,
       volume = {8},
       number = {10},
        pages = {1246-1256},
          doi = {10.1038/s41550-024-02321-9},
       adsurl = {https://ui.adsabs.harvard.edu/abs/2024NatAs...8.1246H}
}

@ARTICLE{Riley2006,
       author = {{Riley}, Pete and {Linker}, J.~A. and {Miki{\'c}}, Z. and {Lionello}, R. and {Ledvina}, S.~A. and {Luhmann}, J.~G.},
        title = "{A Comparison between Global Solar Magnetohydrodynamic and Potential Field Source Surface Model Results}",
      journal = {\apj},
         year = 2006,
        month = dec,
       volume = {653},
       number = {2},
        pages = {1510-1516},
          doi = {10.1086/508565},
       adsurl = {https://ui.adsabs.harvard.edu/abs/2006ApJ...653.1510R}
}

@ARTICLE{Levine1982,
       author = {{Levine}, R.~H. and {Schulz}, M. and {Frazier}, E.~N.},
        title = "{Simulation of the Magnetic Structure of the Inner Heliosphere by Means of a Non-Spherical Source Surface}",
      journal = {\solphys},
         year = 1982,
        month = apr,
       volume = {77},
       number = {1-2},
        pages = {363-392},
          doi = {10.1007/BF00156118},
       adsurl = {https://ui.adsabs.harvard.edu/abs/1982SoPh...77..363L}
}

@ARTICLE{Linker2017,
       author = {{Linker}, J.~A. and {Caplan}, R.~M. and {Downs}, C. and {Riley}, P. and {Mikic}, Z. and {Lionello}, R. and {Henney}, C.~J. and {Arge}, C.~N. and {Liu}, Y. and {Derosa}, M.~L. and {Yeates}, A. and {Owens}, M.~J.},
        title = "{The Open Flux Problem}",
      journal = {\apj},
         year = 2017,
        month = oct,
       volume = {848},
       number = {1},
          eid = {70},
        pages = {70},
          doi = {10.3847/1538-4357/aa8a70},
archivePrefix = {arXiv},
       eprint = {1708.02342},
 primaryClass = {astro-ph.SR},
       adsurl = {https://ui.adsabs.harvard.edu/abs/2017ApJ...848...70L}
}

@ARTICLE{Riley2019,
       author = {{Riley}, Pete and {Linker}, Jon A. and {Mikic}, Zoran and {Caplan}, Ronald M. and {Downs}, Cooper and {Thumm}, Jean-Luc},
        title = "{Can an Unobserved Concentration of Magnetic Flux Above the Poles of the Sun Resolve the Open Flux Problem?}",
      journal = {\apj},
         year = 2019,
        month = oct,
       volume = {884},
       number = {1},
          eid = {18},
        pages = {18},
          doi = {10.3847/1538-4357/ab3a98},
       adsurl = {https://ui.adsabs.harvard.edu/abs/2019ApJ...884...18R}
}

@ARTICLE{Pontin&Wyper2015,
       author = {{Pontin}, D.~I. and {Wyper}, P.~F.},
        title = "{The Effect of Reconnection on the Structure of the Sun's Open-Closed Flux Boundary}",
      journal = {\apj},
         year = 2015,
        month = may,
       volume = {805},
       number = {1},
          eid = {39},
        pages = {39},
          doi = {10.1088/0004-637X/805/1/39},
archivePrefix = {arXiv},
       eprint = {1502.01311},
 primaryClass = {astro-ph.SR},
       adsurl = {https://ui.adsabs.harvard.edu/abs/2015ApJ...805...39P}
}

@ARTICLE{Antiochos2011,
       author = {{Antiochos}, S.~K. and {Miki{\'c}}, Z. and {Titov}, V.~S. and {Lionello}, R. and {Linker}, J.~A.},
        title = "{A Model for the Sources of the Slow Solar Wind}",
      journal = {\apj},
         year = 2011,
        month = apr,
       volume = {731},
       number = {2},
          eid = {112},
        pages = {112},
          doi = {10.1088/0004-637X/731/2/112},
archivePrefix = {arXiv},
       eprint = {1102.3704},
 primaryClass = {astro-ph.SR},
       adsurl = {https://ui.adsabs.harvard.edu/abs/2011ApJ...731..112A}
}

@ARTICLE{Lockwood2022,
       author = {{Lockwood}, Mike and {Owens}, Mathew J. and {Yardley}, Stephanie L. and {Virtanen}, Iiro O.~I. and {Yeates}, Anthony R. and {Mu{\~n}oz-Jaramillo}, Andr{\'e}s},
        title = "{Application of historic datasets to understanding open solar flux and the 20th-century grand solar maximum. 2. Solar observations}",
      journal = {Frontiers in Astronomy and Space Sciences},
         year = 2022,
        month = sep,
       volume = {9},
          eid = {976444},
        pages = {976444},
          doi = {10.3389/fspas.2022.976444},
       adsurl = {https://ui.adsabs.harvard.edu/abs/2022FrASS...9.6444L}
}

@ARTICLE{Panasenco2020,
       author = {{Panasenco}, Olga and {Velli}, Marco and {D'Amicis}, Raffaella and {Shi}, Chen and {R{\'e}ville}, Victor and {Bale}, Stuart D. and {Badman}, Samuel T. and {Kasper}, Justin and {Korreck}, Kelly and {Bonnell}, J.~W. and {Wit}, Dudok de Thierry and {Goetz}, Keith and {Harvey}, Peter R. and {MacDowall}, Robert J. and {Malaspina}, David M. and {Pulupa}, Marc and {Case}, Anthony W. and {Larson}, Davin and {Livi}, Roberto and {Stevens}, Michael and {Whittlesey}, Phyllis},
        title = "{Exploring Solar Wind Origins and Connecting Plasma Flows from the Parker Solar Probe to 1 au: Nonspherical Source Surface and Alfv{\'e}nic Fluctuations}",
      journal = {\apjs},
         year = 2020,
        month = feb,
       volume = {246},
       number = {2},
          eid = {54},
        pages = {54},
          doi = {10.3847/1538-4365/ab61f4},
       adsurl = {https://ui.adsabs.harvard.edu/abs/2020ApJS..246...54P}
}

@ARTICLE{Ma2025,
       author = {{Ma}, Xinyi and {Yang}, Liping and {Feng}, Xueshang and {Tian}, Hui and {Wu}, Honghong and {Shen}, Fang and {Zhang}, Wangning and {Ma}, Mengxuan and {Zhang}, Xiao and {Wang}, Ziwei},
        title = "{Backmapping of the High- and Low-latitude Solar Wind under Multiple Heliospheric and Coronal Magnetic Field Configurations}",
      journal = {\apj},
         year = 2025,
        month = dec,
       volume = {994},
       number = {2},
          eid = {160},
        pages = {160},
          doi = {10.3847/1538-4357/ae0f1e},
archivePrefix = {arXiv},
       eprint = {2510.21076},
 primaryClass = {astro-ph.SR},
       adsurl = {https://ui.adsabs.harvard.edu/abs/2025ApJ...994..160M}
}

@ARTICLE{Abbo2016,
       author = {{Abbo}, L. and {Ofman}, L. and {Antiochos}, S.~K. and {Hansteen}, V.~H. and {Harra}, L. and {Ko}, Y.-K. and {Lapenta}, G. and {Li}, B. and {Riley}, P. and {Strachan}, L. and {von Steiger}, R. and {Wang}, Y.-M.},
        title = "{Slow Solar Wind: Observations and Modeling}",
      journal = {\ssr},
         year = 2016,
        month = nov,
       volume = {201},
       number = {1-4},
        pages = {55-108},
          doi = {10.1007/s11214-016-0264-1},
       adsurl = {https://ui.adsabs.harvard.edu/abs/2016SSRv..201...55A}
}

@article{sullivan2019pyvista,
  doi = {10.21105/joss.01450},
  url = {https://doi.org/10.21105/joss.01450},
  year = {2019},
  month = {May},
  publisher = {The Open Journal},
  volume = {4},
  number = {37},
  pages = {1450},
  author = {Bane Sullivan and Alexander Kaszynski},
  title = {{PyVista}: {3D} plotting and mesh analysis through a streamlined interface for the {Visualization Toolkit} ({VTK})},
  journal = {Journal of Open Source Software}
}

@misc{BarattaEtal2023,
  title     = {{DOLFINx}: the next generation {FEniCS} problem solving environment},
  author    = {Baratta, Igor A. and Dean, Joseph P. and Dokken, J{\o}rgen S. and Habera, Michal and Hale, Jack S. and Richardson, Chris N. and Rognes, Marie E. and Scroggs, Matthew W. and Sime, Nathan and Wells, Garth N.},
  doi       = {10.5281/zenodo.10447666},
  year      = {2023},
  howpublished = {preprint}
}

@ARTICLE{Bale2019,
       author = {{Bale}, S.~D. and {Badman}, S.~T. and {Bonnell}, J.~W. and {Bowen}, T.~A. and {Burgess}, D. and {Case}, A.~W. and {Cattell}, C.~A. and {Chandran}, B.~D.~G. and {Chaston}, C.~C. and {Chen}, C.~H.~K. and {Drake}, J.~F. and {de Wit}, T. Dudok and {Eastwood}, J.~P. and {Ergun}, R.~E. and {Farrell}, W.~M. and {Fong}, C. and {Goetz}, K. and {Goldstein}, M. and {Goodrich}, K.~A. and {Harvey}, P.~R. and {Horbury}, T.~S. and {Howes}, G.~G. and {Kasper}, J.~C. and {Kellogg}, P.~J. and {Klimchuk}, J.~A. and {Korreck}, K.~E. and {Krasnoselskikh}, V.~V. and {Krucker}, S. and {Laker}, R. and {Larson}, D.~E. and {MacDowall}, R.~J. and {Maksimovic}, M. and {Malaspina}, D.~M. and {Martinez-Oliveros}, J. and {McComas}, D.~J. and {Meyer-Vernet}, N. and {Moncuquet}, M. and {Mozer}, F.~S. and {Phan}, T.~D. and {Pulupa}, M. and {Raouafi}, N.~E. and {Salem}, C. and {Stansby}, D. and {Stevens}, M. and {Szabo}, A. and {Velli}, M. and {Woolley}, T. and {Wygant}, J.~R.},
        title = "{Highly structured slow solar wind emerging from an equatorial coronal hole}",
      journal = {\nat},
         year = 2019,
        month = dec,
       volume = {576},
       number = {7786},
        pages = {237-242},
          doi = {10.1038/s41586-019-1818-7},
       adsurl = {https://ui.adsabs.harvard.edu/abs/2019Natur.576..237B}
}

@ARTICLE{Higginson2018,
       author = {{Higginson}, A.~K. and {Lynch}, B.~J.},
        title = "{Structured Slow Solar Wind Variability: Streamer-blob Flux Ropes and Torsional Alfv{\'e}n Waves}",
      journal = {\apj},
         year = 2018,
        month = may,
       volume = {859},
       number = {1},
          eid = {6},
        pages = {6},
          doi = {10.3847/1538-4357/aabc08},
archivePrefix = {arXiv},
       eprint = {1710.00106},
 primaryClass = {astro-ph.SR},
       adsurl = {https://ui.adsabs.harvard.edu/abs/2018ApJ...859....6H}
}

@ARTICLE{Zhukov2025,
       author = {{Zhukov}, A.~N. and {Thizy}, C. and {Galano}, D. and {Bourgoignie}, B. and {Dolla}, L. and {Jean}, C. and {Nicula}, B. and {Shestov}, S. and {Galy}, C. and {Rougeot}, R. and {Versluys}, J. and {Zender}, J. and {Lamy}, P. and {Fineschi}, S. and {Gunar}, S. and {Inhester}, B. and {Mierla}, M. and {Rudawy}, P. and {Tsinganos}, K. and {Koutchmy}, S. and {Howard}, R. and {Peter}, H. and {Vives}, S. and {Abbo}, L. and {Aime}, C. and {Aleksiejuk}, K. and {Baran}, J. and {Bak-Steslicka}, U. and {Bemporad}, A. and {Berghmans}, D. and {Besliu-Ionescu}, D. and {Buckley}, S. and {Buiu}, O. and {Capobianco}, G. and {Cimoch}, I. and {DHuys}, E. and {Dziezyc}, M. and {Fleury-Frenette}, K. and {Gibson}, S.~E. and {Giordano}, S. and {Golub}, L. and {Grochowski}, K. and {Heinzel}, P. and {Hermans}, A. and {Jacobs}, J. and {Jejcic}, S. and {Kranitis}, N. and {Landini}, F. and {Loreggia}, D. and {Magdalenic}, J. and {Maia}, D. and {Marque}, C. and {Melich}, R. and {Morawski}, M. and {Mosdorf}, M. and {Noce}, V. and {Orleanski}, P. and {Paschalis}, A. and {Peresty}, R. and {Rodriguez}, L. and {Seaton}, D.~B. and {Short}, L. and {Simar}, J.-F. and {Steslicki}, M. and {Sorensen}, R. and {Terrasa}, G. and {Van Vooren}, N. and {Verstringe}, F. and {Zangrilli}, L.},
        title = "{The ASPIICS solar coronagraph aboard the Proba-3 formation flying mission. Scientific objectives and instrument design}",
      journal = {arXiv e-prints},
         year = 2025,
        month = aug,
          eid = {arXiv:2509.00253},
        pages = {arXiv:2509.00253},
          doi = {10.48550/arXiv.2509.00253},
archivePrefix = {arXiv},
       eprint = {2509.00253},
 primaryClass = {astro-ph.SR},
       adsurl = {https://ui.adsabs.harvard.edu/abs/2025arXiv250900253Z}
}

@ARTICLE{Gmshpaper,
       author = {{Geuzaine}, Christophe and {Remacle}, Jean-Fran{\c{c}}ois},
        title = "{Gmsh: A 3-D finite element mesh generator with built-in pre- and post-processing facilities}",
      journal = {International Journal for Numerical Methods in Engineering},
         year = 2009,
        month = sep,
       volume = {79},
       number = {11},
        pages = {1309-1331},
          doi = {10.1002/nme.2579},
       adsurl = {https://ui.adsabs.harvard.edu/abs/2009IJNME..79.1309G}
}

@ARTICLE{Tahtinen2024,
       author = {{T{\"a}htinen}, Ismo and {Asikainen}, Timo and {Mursula}, Kalevi},
        title = "{Straight outta photosphere: Open solar flux without coronal modeling}",
      journal = {\aap},
         year = 2024,
        month = aug,
       volume = {688},
          eid = {L32},
        pages = {L32},
          doi = {10.1051/0004-6361/202451267},
archivePrefix = {arXiv},
       eprint = {2408.11525},
 primaryClass = {astro-ph.SR},
       adsurl = {https://ui.adsabs.harvard.edu/abs/2024A&A...688L..32T}
}

@ARTICLE{Tahtinen2026,
       author = {{T{\"a}htinen}, Ismo and {Asikainen}, Timo and {Mursula}, Kalevi},
        title = "{Active regions and the large-scale magnetic field of solar cycle 24}",
      journal = {\aap},
         year = 2026,
        month = feb,
       volume = {706},
          eid = {A235},
        pages = {A235},
          doi = {10.1051/0004-6361/202557466},
archivePrefix = {arXiv},
       eprint = {2603.05654},
 primaryClass = {astro-ph.SR},
       adsurl = {https://ui.adsabs.harvard.edu/abs/2026A&A...706A.235T}
}

@ARTICLE{Arge2024,
       author = {{Arge}, C. Nick and {Leisner}, Andrew and {Antiochos}, Spiro K. and {Wallace}, Samantha and {Henney}, Carl J.},
        title = "{Proposed Resolution to the Solar Open Magnetic Flux Problem}",
      journal = {\apj},
         year = 2024,
        month = apr,
       volume = {964},
       number = {2},
          eid = {115},
        pages = {115},
          doi = {10.3847/1538-4357/ad20e2},
archivePrefix = {arXiv},
       eprint = {2304.07649},
 primaryClass = {astro-ph.SR},
       adsurl = {https://ui.adsabs.harvard.edu/abs/2024ApJ...964..115A}
}

@software{Wu2026NSPF,
  author       = {Wu, Ziqi},
  title        = {NSPF: a Non-Spherical Potential Field for coronal
                   magnetic field extrapolations
                  },
  month        = apr,
  year         = 2026,
  publisher    = {Zenodo},
  version      = {v1.0},
  doi          = {10.5281/zenodo.19921187},
  url          = {https://doi.org/10.5281/zenodo.19921187},
}

@ARTICLE{Rice2021,
       author = {{Rice}, Oliver E.~K. and {Yeates}, Anthony R.},
        title = "{Global Coronal Equilibria with Solar Wind Outflow}",
      journal = {\apj},
         year = 2021,
        month = dec,
       volume = {923},
       number = {1},
          eid = {57},
        pages = {57},
          doi = {10.3847/1538-4357/ac2c71},
archivePrefix = {arXiv},
       eprint = {2110.01319},
 primaryClass = {astro-ph.SR},
       adsurl = {https://ui.adsabs.harvard.edu/abs/2021ApJ...923...57R}
}

@ARTICLE{Heinemann2026,
       author = {{Heinemann}, Stephan G. and {Pomoell}, Jens and {Temmer}, Manuela},
        title = "{Magnetic Topology and Loop Statistics in Observed Coronal Holes Using Potential Field Modeling}",
      journal = {\apj},
         year = 2026,
        month = feb,
       volume = {998},
       number = {1},
          eid = {176},
        pages = {176},
          doi = {10.3847/1538-4357/ae39c4},
archivePrefix = {arXiv},
       eprint = {2601.11080},
 primaryClass = {astro-ph.SR},
       adsurl = {https://ui.adsabs.harvard.edu/abs/2026ApJ...998..176H}
}
\bibliographystyle{aasjournalv7}
\end{document}